\documentclass[fleqn,usenatbib]{mnras}

\usepackage[T1]{fontenc}
\usepackage{graphicx}
\usepackage{xcolor}
\usepackage{color}
\usepackage{hyperref}
\usepackage{notes2bib}
\usepackage{float}
\usepackage{array}
\usepackage{graphicx} 
\usepackage{multirow}
\usepackage{hhline,booktabs}
\usepackage{makecell}
\usepackage{amsmath,amssymb}
\usepackage{newtxtext,newtxmath}
\hypersetup{colorlinks=true,linkcolor=[rgb]{1.,0.2,0.2},citecolor=[rgb]{0.1,0.4,1.},filecolor=[rgb]{0.7,0.2,0.2},urlcolor=[rgb]{0.0,0.2,1.}}
\makeatletter
\@fleqnfalse
\@mathmargin\@centering
\makeatother

\def\INSPIRE{\mbox{{\tt INSPIRE}}}
\def\EINSPIRE{\mbox{\tt E-INSPIRE}}
\newcommand{\Reff}{$\mathrm{R}_{\mathrm{e}\,}$}
\newcommand{\Mstar}{$\mathrm{M}_{\star}\,$}
\newcommand{\Mfrac}{$f_{M^{\star}_{t\text{BB}=3}}$}

\newcommand{\afe}{[$\alpha$/Fe]}
\newcommand{\kms}{km s$^{-1}$}
\newcommand{\Msun}{M$_{\odot}\,$}
\usepackage[normalem]{ulem}
\newcommand{\ppxf}{\textsc{pPXF}}

\definecolor{darkgreen}{rgb}{0.09, 0.45, 0.27}
\definecolor{amber}{rgb}{1.0, 0.49, 0.0}

\defcitealias{Ferre-Mateu+17}{F17}
\defcitealias{Tortora+18_UCMGs}{T18}
\defcitealias{Scognamiglio20}{S20}
\defcitealias{Spiniello20_Pilot}{\INSPIRE\, Pilot}
\defcitealias{Spiniello+21}{\INSPIRE\, DR1}
\defcitealias{DAgo23}{\INSPIRE\, DR2}
\defcitealias{Spiniello24}{\INSPIRE\, DR3}
\title[A statistical sample of UCMGs at $z<0.3$]{\centering \EINSPIRE\ - I. 
Bridging the gap with the local Universe: Stellar population of a statistical sample of ultra-compact massive galaxies at $z<0.3$}
\author[J.Mills et al.]{\noindent
John Mills$^{1}$, Chiara Spiniello$^{1, 2}$\thanks{E-mail: chiara.spiniello@physics.ox.ac.uk},
Alexey Sergeyev$^{3,4,5}$, Crescenzo Tortora$^{2}$,
Vladyslav Khramtsov$^{6}$,  \and
Giuseppe D'Ago$^{7}$, Michalina~Maksymowicz-Maciata$^{8}$, Jo\~ao P. V. Benedetti$^{9, 10}$, Anna Ferr\'e-Mateu$^{9, 10}$, \and
Michele Cappellari$^{1}$, Roger Davies$^{1}$, Johanna Hartke$^{11,12}$, Charles Rosen$^{1}$
\\ 
$^{1}$ Sub-Dep. of Astrophysics, Dep. of Physics, University of Oxford, Denys Wilkinson Building, Keble Road, Oxford OX1 3RH, United Kingdom\\
$^{2}$ INAF -  Osservatorio Astronomico di Capodimonte, Via Moiariello  16, 80131, Naples, Italy\\
$^{3}$ Université Côte d'Azur, Observatoire de la Côte d'Azur, CNRS, Laboratoire Lagrange, France \\
$^{4}$ V.N. Karazin Kharkiv National University, Sumska 35, Ukraine\\
$^{5}$ Institute of Radio Astronomy of National Academy of Science of Ukraine, Mystetstv 4, Ukraine\\
$^{6}$ Department of Astronomy and Space Informatics, V. N. Karazin Kharkiv National University, 35 Sumska Str., Kharkiv, Ukraine \\
$^{7}$ Institute of Astronomy, University of Cambridge, Madingley Road, Cambridge CB3 0HA, United Kingdom\\
$^{8}$ Astrophysics Group, H. H. Wills Physics Laboratory, University of Bristol, Tyndall Avenue, Bristol, BS1 8TL, United Kingdom\\
$^{9}$ Instituto de Astrof\'isica de Canarias, V\'ia L\'actea s/n, E-38205 La Laguna, Tenerife, Spain\\
$^{10}$ Departamento de Astrof\'isica, Universidad de La Laguna, E-38200, La Laguna, Tenerife, Spain\\
$^{11}$ Finnish Centre for Astronomy with ESO (FINCA), FI-20014 University of Turku, Finland\\
$^{12}$ Tuorla Observatory, Department of Physics and Astronomy, University of Turku, 20014 Turku, Finland\\
}
\date{Accepted XXX. Received YYY; in original form ZZZ}

\pubyear{2024}

\begin{document}
\label{firstpage}
\pagerange{\pageref{firstpage}--\pageref{lastpage}}
\maketitle
\begin{abstract}
This paper presents the first effort to Extend the Investigation of Stellar Populations In RElics (\EINSPIRE). We present a catalogue of 430 spectroscopically-confirmed ultra-compact massive galaxies (UCMGs) from the Sloan Digital Sky Survey at redshifts $0.01<z<0.3$. This increases the original \INSPIRE\ sample eightfold, bridging the gap with the local Universe. For each object, we compute integrated stellar velocity dispersion, age,  metallicity, and [Mg/Fe] through spectroscopic stellar population analysis. We infer star formation histories (SFHs), metallicity evolution histories (MEHs) and compute the \textit{Degree of Relicness} (DoR) of each object. The UCMGs, covering a wide range of DoR from 0.05 to 0.88, can be divided into three groups, according to how extreme their SFH was. The first group consists of 81 extreme relics ($\text{DoR}\gtrsim0.6$) that have formed the totality of their stellar mass by $z\sim2$ and have super-solar metallicities at all cosmic epochs. The second group ($0.3\lesssim\text{DoR}\lesssim0.6$) contains 293 objects also characterised by peaked SFHs but with a small percentage of later-formed stars and with a variety of MEHs.  
The third group ($\text{DoR}\lesssim0.3$), has 56 objects that cannot be considered relics since they have extended SFHs and formed a non-negligible fraction ($>25$ \%) of their stellar mass at $z<2$. We confirm that an efficient method of finding relics is to select UCMGs with large velocity dispersion values but we believe that the most efficient way is to select high velocity dispersion objects that also have super-solar metallicities and high [Mg/Fe].
\end{abstract}



\section{Introduction}
Relics \citep{Trujillo+09_superdense}, the nearby counterpart of high-$z$ "red-nuggets" \citep{Damjanov+09}, are the perfect local laboratories to investigate the early phases of the mass assembly and formation scenarios of massive galaxies. In fact, they are almost exclusively composed of "pristine", very old stellar populations that formed at very high-$z$ and evolved passively and undisturbed thereafter. 

Relics are, however, very rare: their local density at $z<0.5$ ranges between $10^{-7}$ and $10^{-5}$ $\text{Mpc}^{-3}$ and declines at low redshift \citep{Ferre-Mateu+17,Charbonnier+17_compact_galaxies,Tortora+18_UCMGs,Spiniello+21}. This is the reason why the INvestigating Stellar Population In RElics (\INSPIRE) Project  started a systematic effort to build a statistically large sample of relics \citep{Spiniello20_Pilot, Spiniello+21, DAgo23, Spiniello24, Martin-Navarro+23, Maksymowicz-Maciata24}. 
Thanks to an ESO Observational Large Program, the \INSPIRE\ survey has targeted, with the X-Shooter spectrograph (XSH, \citealt{Vernet11}), 52 spectroscopically confirmed red ultra-compact massive galaxies (UCMGs). These objects are defined as outliers on the stellar-mass size relation for ETGs \citep[e.g.,][]{Shen+03}, being 4-5 times more compact than normal-sized galaxies with similar stellar mass. Many mass and size thresholds have been used in the literature to define UCMGs \citep{Charbonnier+17_compact_galaxies}. In particular, the 52 \INSPIRE\ targets have been selected to have  optical median effective radii \Reff$<2$ kpc, stellar masses \Mstar$>6\times 10^{10}$ \Msun, redshifts $0.1<z<0.4$, and red colours (see Fig.~1 in \citealt{Spiniello+21}). Among these, 38 had formed 75\% or more of their stellar mass at $z>2$, and were thus classified as relics \citep{Spiniello24}.  
In terms of structural parameters and photometry, both relics and non-relics in \INSPIRE\ are red, but span a large range of 
S\'ersic indices ($n$), and axis ratios ($q$). These quantities were obtained in \citet{Tortora+16_compacts_KiDS, Tortora+18_UCMGs} and \citet{Scognamiglio20} by fitting a point-spread function (PSF) convolved S\'ersic profile to $g,r,i$ images from the Kilo Degree Survey (KiDS, \citealt{Kuijken11}),  using the code \textsc{2dphot} \citep{LaBarbera_08_2DPHOT}.  

For each of the UCMGs, \citet{Spiniello24}\footnote{The 5th paper of the series which also presented the third ESO data release} computed a \textit{Degree of Relicness} (DoR, \citealt{Ferre-Mateu+17}). 
The DoR is a dimensionless number obtained as the mean of three quantities: the fraction of stellar mass formed by $z=2$, the inverse of the cosmic time at which a galaxy has assembled 75\% of its mass, and the inverse of the final assembly time renormalised by the redshifts of the objects (see Eq.~1 in \citetalias{Spiniello24}). 
Hence, practically, a higher DoR corresponds to a very early and quick mass assembly: the most extreme relics that have assembled the totality of its stellar mass only 0.7 Gyr after the Big Bang would have DoR $=1$, while an object with a very extended star formation history (SFH) that is still forming a small percentage of stars would have DoR $=0$. For reference, by this definition NGC~1277, the bonafide massive relic in the local Universe, has a DoR of $\sim0.95$.

Characterising the first large sample of relics, \INSPIRE\ has provided
hints on how to search for these rare objects. Hence, now, with the Extension of the INvestigating  Stellar Populations In RElics (\EINSPIRE) project, we aim at extending in redshift, stellar mass and wavelength the current sample of UCMGs with a measured DoR,  with the goal of collecting a statistically large sample of ultra-compact galaxies, looking for the most extreme relics in our Universe. Since relics are the `building blocks' of today's giant ETGs, the \EINSPIRE\ project will allow us to shed light on the formation and evolution of this population of galaxies that 
account for more than half of
the total stellar mass in the Universe and are responsible for most of its chemical enrichment.

In this paper, the first of the new series, we start by extending the redshift coverage of \INSPIRE\ by searching for relics in the redshift window $0.01<z<0.3$, hence bridging the gap with the very few known relics in the local Universe \citep{Trujillo+14, Ferre-Mateu+17, Yildirim17}. 

One of the main results obtained so far from \INSPIRE\ is that the DoR strongly correlates with the integrated stellar velocity dispersion ($\sigma_{\star}$). Relics (UCMGs with higher DoR) have larger $\sigma_{\star}$ than non-relic UCMGs and normal-sized ETGs of similar stellar mass \citep{Spiniello+21, Maksymowicz-Maciata24}. Hence, selecting compact galaxies with a very high velocity dispersion seems to be a very effective way to select relic candidates. 
There have been previous efforts to search for compact galaxies with higher stellar velocity dispersions at low redshifts within the Sloan Digital Sky Survey (SDSS), such as \citet{Saulder+15_compacts} and \citet{Clerici+24}. SDSS is ideal to search for these very rare systems due to the wide sky coverage of more than 8000 square degrees. Very recently, \citet{Clerici+24} presented a stellar population study of 1858 compact massive galaxies selected to be outliers in the $\log \sigma_{\star}$ - $\log$ \Reff and in the $\log M_{\star}$ - $\log \sigma_{\star}$ planes. However, their sample was limited to $\sigma_{\star}<380$ \kms\ due to the belief that values above this were not to be trusted. While this may be true in some cases, previous results have demonstrated that both nearby relics \citep{Trujillo+14, Ferre-Mateu+17, Spiniello24} and high-z red nuggets \citep{Saracco20} can have velocity dispersions even higher than this value. Furthermore, an incredibly high value measured for the integrated velocity dispersion could also be due to a combination of an unresolved rotation added on top of a large velocity dispersion with a rather flat spatial profile. Hence, we believe that  the most effective approach to search for relics requires searching through the full range of $\sigma_{\star}$, and then manually checking the full spectral fitting to assess its quality. This is what we proceeded to do in this paper. 

The paper is organised as follows. In Section~\ref{sec:data} we introduce the sample, highlighting the selection criteria used to define UCMGs. In Section~\ref{sec:analysis} we present the spectral analysis, including the computation of the signal-to-noise ratio (SNR), the integrated stellar velocity dispersion ($\sigma_{\star}$), and the stellar population parameters. In particular, we infer the SSP-equivalent [Mg/Fe] ratios with line-indices and  age and metallicity via full-spectral fitting. Section~\ref{sec:results} presents our main findings: star formation history and metallicity evolution, as well as the calculation of the DoR for each of the objects and the subsequent confirmation of 435 new relics at $z\le0.3$.  We finally conclude in Section~\ref{sec:conclusions}.

Throughout the paper, we assume a standard $\Lambda$CDM cosmology with $H_0=67.7$ \kms Mpc$^{-1}$, $\Omega_{\mathrm{\Lambda}}=0.689$, and $\Omega_{\mathrm{M}}=0.311$ \citep{Planck+20}.

\section{Data selection}
\label{sec:data}
This section outlines the steps we undertake to identify compact galaxies and select UCMGs from optical  Sloan Digital Sky Survey (SDSS) images and associated, publicly available photometric catalogues. 

\subsection{Initial selection of objects} 
\label{sec:selection}
We use spectroscopic data from the SDSS DR18 \citep{Almeida+23_SDSS} to select extra-galactic, bright, and red objects with high stellar velocity dispersion values. 
In particular, we run a SQL search retrieving objects from the imaging (PhotoObj) and spectroscopic (SpecObj) SDSS catalogues with the following criteria:
\begin{itemize}
    \item redshift $0.005<z<0.4$. The lower limit ensures the removal of galactic contaminants (stars), while the upper limit is set to have the same stellar absorption lines available for \INSPIRE\ objects;   
    \item Petrosian magnitude in the $r-$band mag$_r<20$,which roughly corresponds to the same magnitude cut of \INSPIRE objects and makes the recovery of structural parameters trustable; 
    \item colour $g-i>1.2$, to select systems with an evolved stellar populations;
    \item stellar velocity dispersion $\sigma_{\star}>200$ \kms, which is the typical values for normal-sized galaxies with \Mstar$\ge10^{10}$\Msun. However, we exclude objects with $\sigma_{\star}=850$ \kms\ exactly (the maximum value given in the catalogues) because this is likely more indicative of a bad fit rather than an object with a very high $\sigma_{\star}$.
\end{itemize}

Furthermore, we only keep systems with no `zWarning' and exclude those with the keyword `class' set to `STARBURST', `STARFORMING', `STARFORMING BROADLINE', or `STARBURST BROADLINE'. As stated in \cite{Bolton+12_SDSSIII}, these classes are defined as systems with emission lines in H$\beta$, [OIII] 5007, H$\alpha$, and [NII] 6583 detected at the 3$\sigma$ level which also satisfy $\log_{10}([\rm OIII]/H\beta)<1.2\,\log_{10}([NII]/H\alpha)+0.22$. Systems are further designated `STARFORMING' if the equivalent width of the H$\alpha$ line is less than 50\AA\ and `STARBURST' if it is greater than 50\AA. A system is set as `BROADLINE' if the line widths are greater than 200\kms with line-width measurement at the 5$\sigma$ level and line-slux measurement at the 10$\sigma$ level. Since  relics, by definition, have formed most of their stellar mass by $z=2$, they should have low star-formation rates. The described SQL search produces an initial shortlist of 387,949 objects. 

To find objects that are UCMGs, we need to obtain an estimate of their stellar masses (\Mstar). We therefore cross-match our shortlist with the 
GALEX-SDSS-WISE Legacy Catalogue, specifically GSWLC-2 \citep{Salim+18}. 
The catalogue 
contains stellar masses, dust attenuation and star formation rates of $\sim700,000$ galaxies with SDSS redshifts below 0.3 and magnitudes mag$_{r}<18$. 
The galaxy properties are obtained via spectral energy distribution (SED) fitting using a Bayesian framework on joint UV+optical+mid-IR data. 
From these quantities, we also calculate the specific star-formation rate (sSFR) for each object, simply dividing the star formation rate by the stellar mass. 
Cross-matching the GSWLC-2 catalogue with our initial shortlist returns 
126,898 objects, where we exclude any objects with `flag\_sed' $\neq0$ which indicates a poor fit. We also note that this cross-match introduces an additional cut in both redshift and $r-$band magnitude, since GSWLC-2 only covers $0.01<z<0.3$ and mag$_r<18.0$. 

\subsection{Selection of compact galaxies using compactness}
\label{sec:compact}
SDSS provides several proxies for the effective radii (\Reff) of the galaxies. In particular, we consider the estimates 
from the de Vaucouleurs (deVRad) and exponential (expRad) fits, along with the radii containing 50\% of the Petrosian flux (petroR50). 
The Petrosian flux is defined in SDSS as the total flux within two Petrosian radii, where the Petrosian radius (petroRad) is the radius at which the ratio of the local surface brightness in an annulus to the mean surface brightness within the same radius reaches the value of 0.2 \citep{Blanton01, Yasuda01, Stoughton+02}. Even though the Petrosian radius is the least dependent on redshift\footnote{According to \citet{Blanton01} the flux contained in the Petrosian half-light radius is completely independent of redshift except when the size of the galaxy is comaparable to the seeing, which is the case for our data.}, we cannot select UCMGs directly on the basis of this quantity since it does not take into account the point spread function (PSF).  
This has a noticeable effect for objects with \Reff$<2$ arcsec \citep{Blanton01}, which includes many of our objects of interest.

SDSS also provides a coefficient ranging between 0 and 1 called FRACDEV ($f_{\mathrm{deV}}$) which describes 
the relative contribution to the total luminosity of the de Vaucouleurs fit with respect to the exponential one \citep{Abazajian+04}. 

To estimate the sizes of each object, we follow the approximation of \cite{Baldry21} and use the geometric mean of the two models, weighted by the $f_{\mathrm{deV}}$ coefficient\footnote{This measure of \Reff compares well with the \cite{Simard+11} catalogue \citep{Baldry21}, especially when using the geometric mean as opposed to the linear weighted mean}. 
: 
\begin{equation}
    \log \mathrm{R_e} = f_\mathrm{{deV}}\log \mathrm{R_{deV}} + (1-f_{\mathrm{deV}})\log \mathrm{R_{exp}}.
\end{equation}

Finally, we convert the effective radius of every object from arcsec to kpc with the \textsc{astropy.cosmo} python package \citep{Astropy22}\footnote{\url{https://www.astropy.org}}, where we use the redshifts obtained from SDSS.

\begin{figure}
    \centering
\includegraphics[width=\columnwidth]{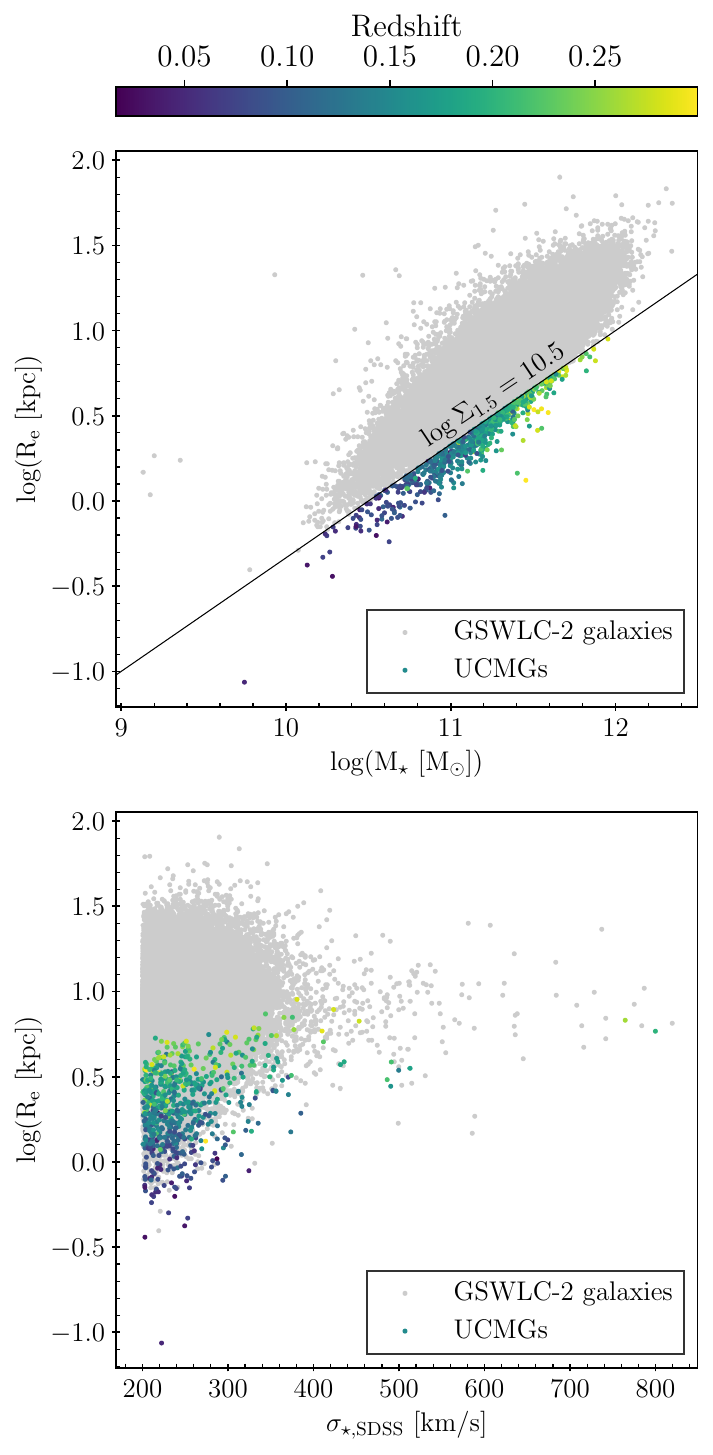}
    \caption{The relations between size and stellar mass (top) and size and velocity dispersion (bottom) for objects in our sample (coloured points) and the entire GSWLC-2 catalogue (grey points). We have selected as UCMGs the objects that satisfy $\log\Sigma_{1.5}=\log($\Mstar$/$\Reff$^{1.5})>10.5$. The UCMGs are colour-coded by redshift. We observe a positive correlation between redshift and stellar mass (and therefore effective radius) because we are only concerned with objects with mag$_r<18$, which systematically filters out less massive objects which are further away.}
    \label{fig:compact}
\end{figure}

There are many criteria by which one can select UCMGs. The most commonly used in the literature are those setting two separate thresholds on the stellar mass and the size (see e.g. \citealt{Charbonnier+17_compact_galaxies} for a list of commonly used definitions). Previously in \INSPIRE\ we have followed this approach and  defined a UCMG as an object with \Reff$<2$ kpc and \Mstar$>6\times 10^{10}$ \Msun \citep{Spiniello+21}. The main limitation of this criterion, however, is that it neglects the size-mass distribution of galaxies, plotted in the top panel of Figure~\ref{fig:compact}. This means that one automatically excludes the most massive objects, which are relatively bigger by definition but still clear outliers with respect to the main population. Here, following \cite{Barro+13}, we define the quantity $\Sigma_{1.5}=$\Mstar$/$\Reff$^{1.5}$ which measures the compactness of the objects. This choice of exponent is roughly consistent with what is found by \cite{Newman+12} for quiescent galaxies and also traces the size-mass distribution observed in Figure \ref{fig:compact}.
Compact galaxies are then defined as outliers in the the size-mass distribution, 
using the same threshold defined in \cite{Baldry21}: $\log\Sigma_{1.5}>10.5$.  


\begin{figure*}
    \centering    \includegraphics[width=0.95\textwidth]{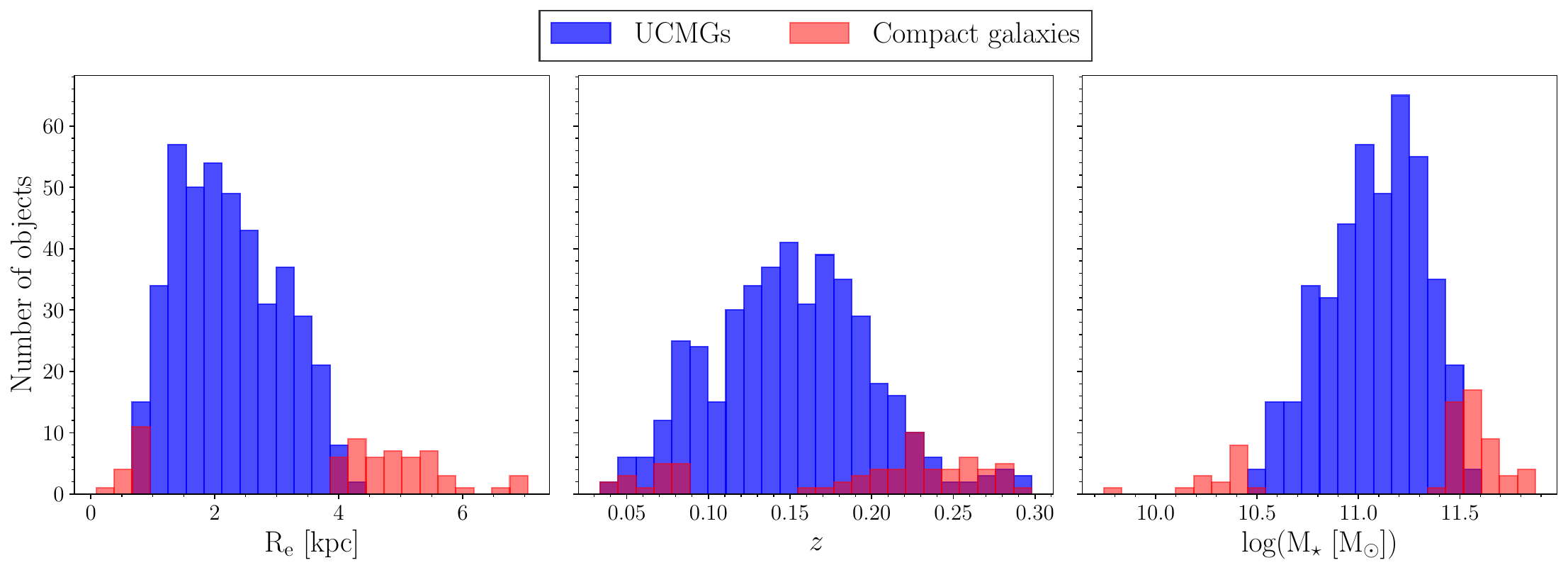}    \caption{Histograms of the effective radii, redshifts, and stellar masses of the 495 compact galaxies. Out of these, 430 can be considered ultra-compact and massive according to the \INSPIRE\ thresholds and are shown in blue. The other compact galaxies (either larger or less massive, as described in Section~\ref{sec:ucmg}) are shown in red and have been excluded from the analysis. Note that the splitting between UCMGs and the rest is not a single threshold in size, but it is obtained using the equation \ref{eq:size} to correct SDSS sizes to the KiDS resolution and then cutting everything with $\mathrm{R}_{\mathrm{e, KiDS}}\le2$ kpc. } 
    \label{fig:hist}
\end{figure*}


In the bottom panel of Figure~\ref{fig:compact}, we plot the effective radii against the velocity dispersion values as computed by SDSS. Compact galaxies  follow a similar distribution as that of the general population of galaxies from the GSWLC-2 catalogue,  
the only difference being that they have smaller \Reff (by definition).

From the galaxies in our input sample, 628 pass the compactness threshold. 
However, we note that among these, a number of contaminants are expected \citep[e.g.,][]{Khramtsov19}. First of all, the classification could be wrong for some of the objects, which can be quasars. 
Or, more commonly, a larger galaxy with a bright stellar nucleus could have a small measured half-right radius as this is set by the nucleus alone. 
Hence, we visually inspected the stamps and SDSS spectra for all the retrieved objects, finally selecting 495 compact galaxies, for which we retrieved the 1D spectra directly from SDSS. Histograms characterising the overall properties of the UCMGs can be found in Figure~\ref{fig:hist}.

\subsection{Selection of UCMGs}
\label{sec:ucmg}
The leftmost histogram of Figure~\ref{fig:hist} shows the effective radii of the 495 compact galaxies. Although the peak of the distribution is at \Reff$<2$ kpc, an extended tail is visible at larger sizes. This is because, differently from what has been done previously, 
we do not impose any upper limit on the effective radii during the selection. 
Hence, this novel approach could, in principle, make it slightly more complicated to perform a direct comparison between this work and previous \INSPIRE\ selections. 
However, we note that the effective radii measured from KiDS data (in \citealt{Tortora+18_UCMGs} and \citealt{Scognamiglio20}), fitting a PSF convolved S\'ersic profile to the images using the code \textsc{2dphot} \citep{LaBarbera_08_2DPHOT}, are systematically smaller than the ones inferred from SDSS imaging. We believe this is because SDSS has a worse spatial resolution and a larger pixel size, as shown in Figure~\ref{fig:kids_compare} of Appendix~\ref{app:size_compare}. There, we compute a correction factor based on the 36 objects that appear in both SDSS and \INSPIRE. When correcting the SDSS-based effective radii, 446 out of the 495 galaxies selected with the compactness criterion have \Reff$_{\rm KiDS}\le 2$ kpc (corresponding to a \Reff$_{\rm SDSS}\lesssim 4$ kpc, see Fig.~\ref{fig:radii_compare}) and hence can be considered ultra-compact 
We further note that our SDSS sample covers a wider range in stellar mass than \INSPIRE, also including a small number of less massive objects. Hence, we further remove from the sample galaxies with \Mstar$\le6\times10^{10}$\Msun. 

In the remainder of the paper we will limit ourselves to the remaining 430 objects which can be considered ultra-compact massive galaxies (UCMGs, blue histograms in Figure~\ref{fig:hist}) and hence compared with the \INSPIRE\ higher-z counterparts. 


\begin{figure}
    \centering    \includegraphics[width=0.98\columnwidth]{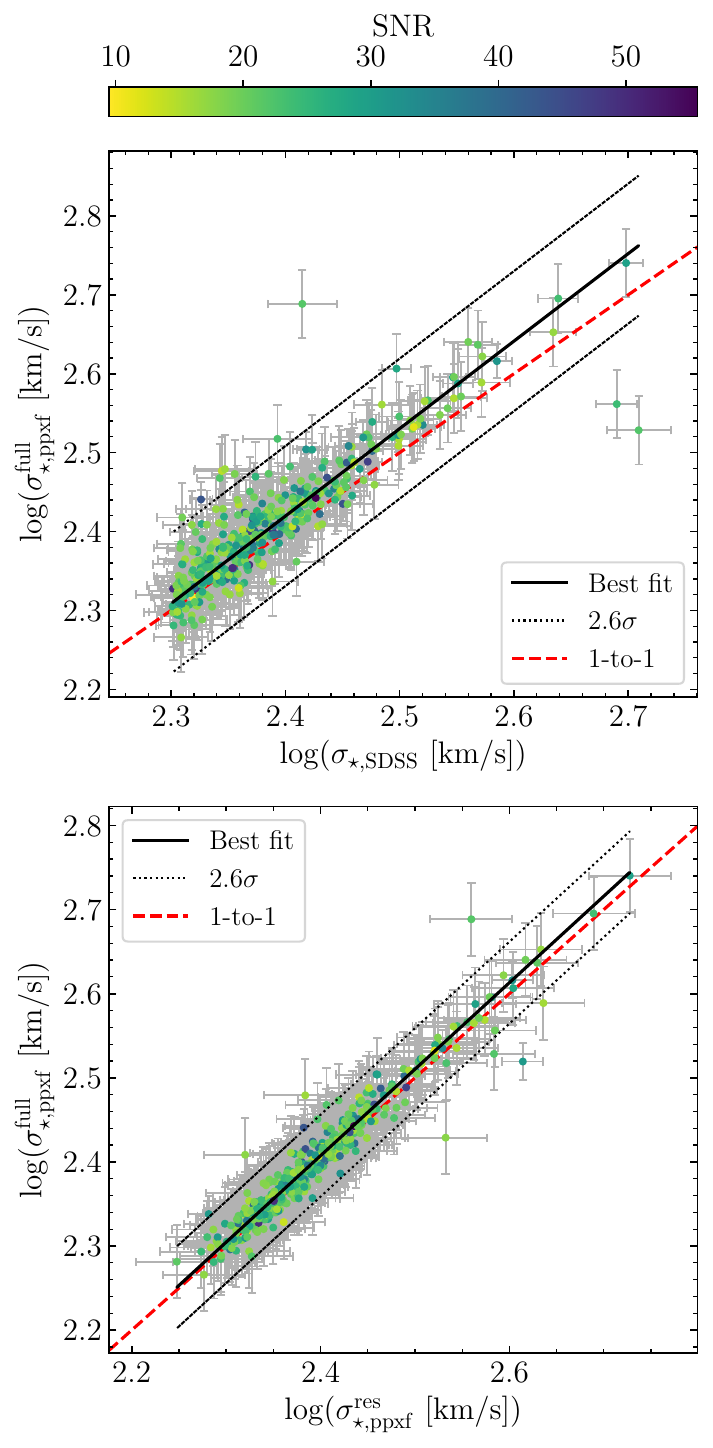}
\caption{\textit{Top panel:} Comparison between the velocity dispersion values retrieved from SDSS (x-axis) and these computed here from the entire wavelength range covered by the spectra.  A good agreement (within 2.6$\sigma$, dotted lines) is found for all but ten objects (see text for more details). \textit{Bottom panel:} Comparison between the velocity dispersion values computed using two different wavelength ranges. A good agreement is found for all but ten objects (see text for more details). For both panels, the best fit line is indicated by a solid black line and the 1-to-1 relation is indicated by a dashed red line. Additionally, the points are colour-coded by the SNR of the spectra.}
    \label{fig:sigmas}
\end{figure}

\begin{figure}
    \centering    \includegraphics[width=0.98\columnwidth]{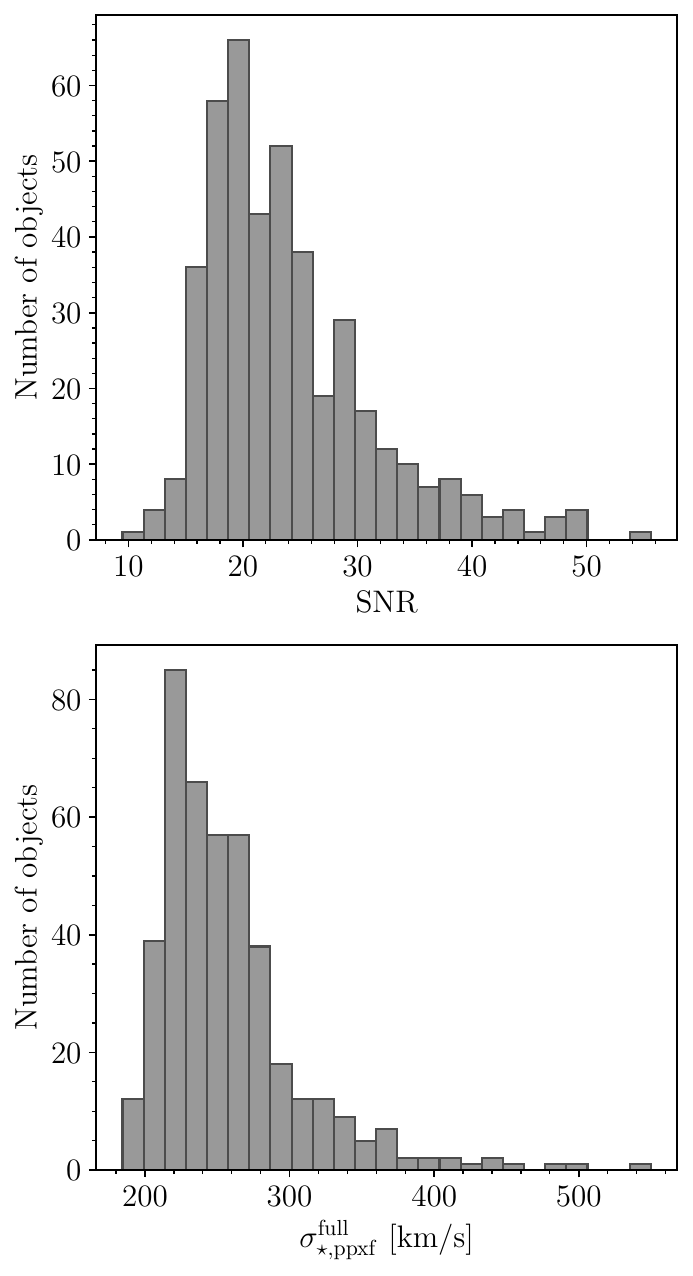}
    \caption{\textit{Top panel: }Histogram of the SNR per dispersive element for the UCMGs. Note that the SNR is calculated over the wavelength range [3600-6500]\AA\ of each spectrum. \textit{Bottom panel: }Histogram of the velocity dispersions of the UCMGs, the calculation of which is discussed in Section~\ref{sec:kinem}.}
    \label{fig:snr}
\end{figure}

\begin{figure}
    \centering    \includegraphics[width=0.95\columnwidth]{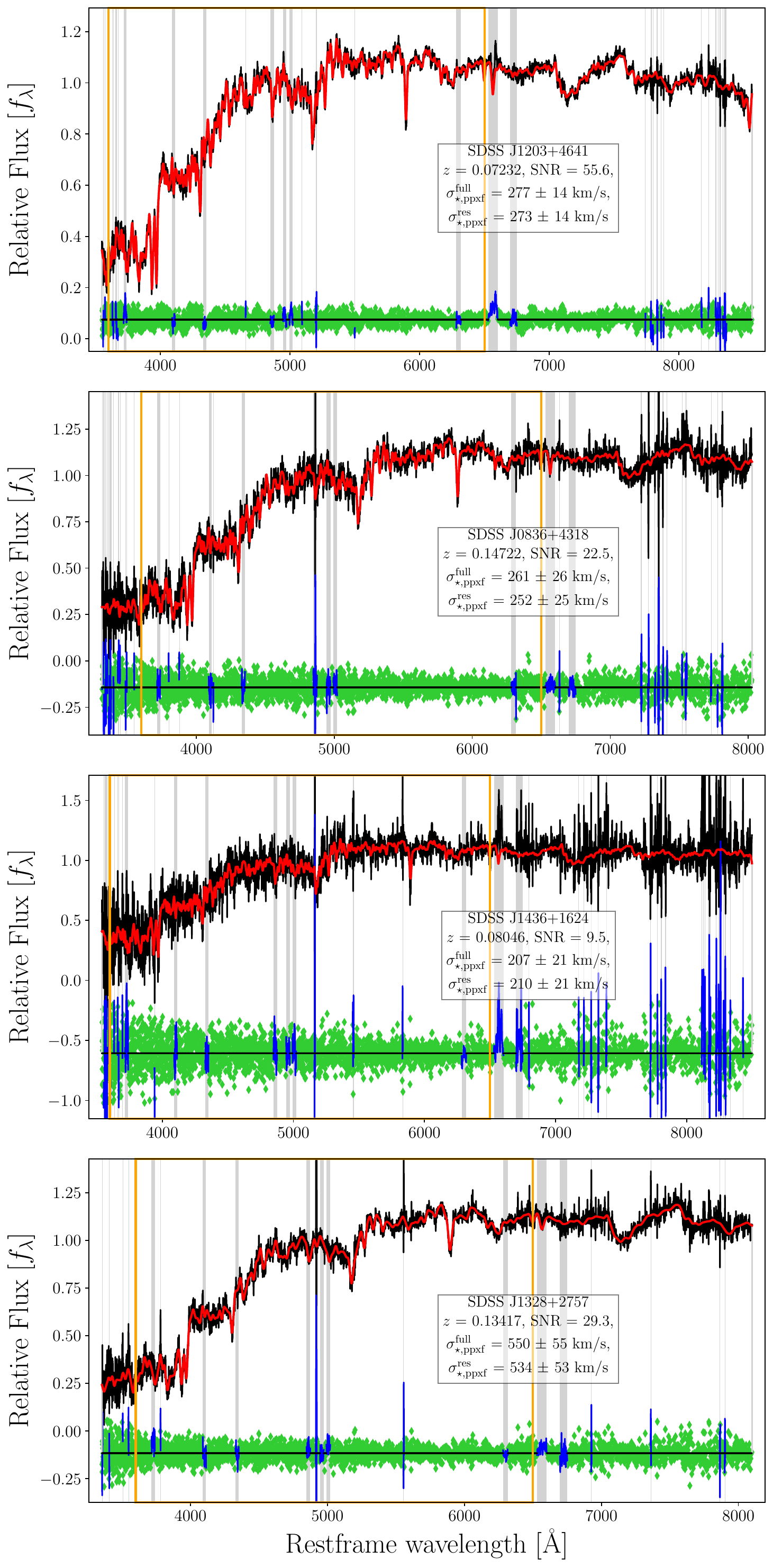}
    \caption{Examples of the \ppxf\ kinematical fit for the (from top to bottom) best, middle, and worst SNR spectra, followed by the spectrum with the highest derived value of $\sigma_{\star}$. The galaxy is plotted in black while the best-fit stellar template is overplotted in red. Green dots are the residuals and blue lines are points excluded by the fit (using the keyword CLEAN). The orange box indicates the wavelength range [3600-6500]\AA, which is the restricted range where both the SNR and $\sigma_{\star, \rm ppxf}^\mathrm{res}$ are calculated. The inset in each panel lists the galaxy ID, the redshift, the integrated SNR and the two resulting velocity dispersions.}
    \label{fig:ppxf}
\end{figure}

\section{Data Analysis}
\label{sec:analysis}

\subsection{Stellar kinematics}
\label{sec:kinem}
Although SDSS has already computed and published the stellar velocity dispersion values for all these systems, we re-derive them using the Penalised Pixel-fitting software\footnote{\url{https://pypi.org/project/ppxf/}} (\ppxf; \citealt{Cappellari04,Cappellari17,Cappellari23}). 
As stellar library for the fit, we employ the E-MILES Single Stellar Population (SSP) models \citep{Vazdekis16}, which are distributed together with the \ppxf\ code. 

For each system, we derive two values of the velocity dispersion, one using the restricted (restframe) wavelength range [3600-6500]\AA\ ($\sigma_{\star, \mathrm{ppxf}}^\mathrm{res}$) for all galaxies, and 
one using the full wavelength range of the spectra ($\sigma_{\star, \mathrm{ppxf}}^\mathrm{full}$). For this latter value the covered wavelength of SDSS ([3800-9200]\AA)  slightly changes from galaxy to galaxy, according to their redshift. 

We compute $\sigma_{\star, \mathrm{ppxf}}^\mathrm{res}$ to be consistent with deriving the stellar population parameters in Section~\ref{sec:stel_pop_ppxf} and to use only the cleanest spectra region. We compute $\sigma_{\star, \mathrm{ppxf}}^\mathrm{full}$ instead  
to maximise the data used to calculate the velocity dispersion and to compare the results we obtain with those published by the SDSS survey. We also note that, given the different redshifts of the objects, the $\sigma_{\star, \mathrm{ppxf}}^\mathrm{full}$ is computed on a slightly different wavelength range for each galaxy. In both cases, we set the additive Legendre polynomial degree (DEGREE) to 20 and use the keyword CLEAN to perform sigma-clipping on the spectra and clean them from residual bad pixels. 
We refer the reader to \citetalias{DAgo23} for extensive testing on how the inferred velocity dispersion values change when changing the parameters of the fit (stellar templates, wavelength range, additive polynomial, masked regions, etc.). In the same paper, we also demonstrated that the systematic uncertainties associated with the velocity dispersion depend on the SNR of the spectrum (see Figure~6 in \citetalias{DAgo23}) and dominate the error budget.  
Hence, following the same approach, although the random errors given by the code are always of the order of 1\%, we attribute a final total (systematic and random) uncertainty of 5\% to the objects with a spectrum with $\rm SNR>30$, and an uncertainty of 10\% for galaxies with spectra of lower SNR. We note that the uncertainties from SDSS only include random errors. 


Overall, a good agreement is found between the velocity dispersion values from SDSS and these inferred by us, as shown in the upper panel of Figure~\ref{fig:sigmas}. The solid line shows the line of best fit, calculated with the \textsc{ltsfit python} package\footnote{\url{https://pypi.org/project/ltsfit/}}, which implements the method described in Section 3.2 of \cite{Cappellari+13_ATLAS3D_XV} and uses the Least Trimmed Squares (LTS, \citealt{LTS}) technique to iteratively clip outliers. The fit is very similar to the 1-to-1 relation (in red), especially for $\sigma_{\star}$ up to $\sim$500 \kms. 
The sigma estimated from \ppxf\ differs by more than 2.6$\sigma$ (containing 99\% of systems if the residuals followed a perfectly Gaussian distribution, dotted lines) only for 10 systems. This means that we would expects roughly 5 spurious outliers out of 500 measures, similar to what we find.  
It should be noted that all but one of these objects have an SNR between $\sim14$ and $\sim24$, which is not particularly high when compared with the overall distribution in Figure~\ref{fig:snr}. The other object has $\rm SNR=43.3$. 

We also observe excellent agreement between the two values of $\sigma_{\star,\mathrm{ppxf}}$ for the vast majority of systems, as shown in the lower panel of Figure~\ref{fig:sigmas}. Similarly to the upper panel, the best fit line is computed by the \textsc{ltsfit python} package and is shown as a solid black line. This fit agrees very well with the 1-to-1 relation (in red) throughout the entire range of $\sigma_{\star}$. Only $10$ out of 430 systems are not within 2.6$\sigma$ of the best fit, however all but three have a SNR between $\sim$16 and $\sim$22, which is not particularly high (see Figure~\ref{fig:snr}). The other three systems have SNRs of 30.6, 30.7, and 43.3. 

Since all three values of velocity dispersion that we have discussed agree very well with each other, we are free to only consider one of them for the rest of the analysis. Henceforth, we shall be referring to $\sigma_{\star, \mathrm{ppxf}}^{\mathrm{full}}$ when referring to velocity dispersion. A histogram of $\sigma_{\star, \mathrm{ppxf}}^{\mathrm{full}}$ for the UCMGs can be found in the bottom panel of Figure~\ref{fig:snr}.

Three examples of the systems with the highest, median and lowest SNR are shown in Figure~\ref{fig:ppxf}, along with the system with the most extreme velocity dispersion as calculated by \ppxf, while the results of SNR and velocity dispersion for a representative sample of the systems are listed in Table~\ref{tab:veldisp}. 
The kinematical results for all 430 galaxies are provided in the online master catalogue associated with this publication\footnote{The full catalogue is available on the \EINSPIRE\ \href{https://sites.google.com/inaf.it/chiara-spiniello/e-inspire}{webpage}}. 

\subsection{Signal-to-noise} 
\label{sec:snr}
We use the python routine \textsc{der\_snr}\footnote{\url{https://www.stecf.org/software/ASTROsoft/DER_SNR/}} to calculate the integrated signal-to-noise (SNR) of each spectrum. This code assumes that the noise is Gaussian and uncorrelated in wavelength bins spaced two pixels apart. We only compute the SNR in the wavelength range in which we fit the stellar population parameters, which is [3600-6500]\AA\ (see section \ref{sec:stel_pop_ppxf}).
The SNR per dispersive element ranges from $\sim$10 to $\sim$56, as shown in the histogram in the upper panel of Figure~\ref{fig:snr}. Such SNRs, listed in Table~\ref{tab:veldisp}, are enough to infer stellar population parameters \citep{Costantin19}.

\begin{table*}
    \centering
\begin{tabular}{lcccccrrr}
\hline
  \multicolumn{1}{c}{ID} &
  \multicolumn{1}{c}{$z$} &
  \multicolumn{1}{c}{SNR} &
  \multicolumn{1}{c}{$\sigma_{\star,\rm SDSS}$} &
  \multicolumn{1}{c}{$\sigma_{\star,\rm ppxf}^{\rm full}$} &
  \multicolumn{1}{c}{$\sigma_{\star,\rm ppxf}^{\rm res}$} &
  \multicolumn{1}{c}{$\log($\Mstar$[$M$_{\odot}])$} &
  \multicolumn{1}{c}{\Reff} &
  \multicolumn{1}{c}{\Reff}\\
  \multicolumn{1}{c}{SDSS} &
  \multicolumn{1}{c}{} &
  \multicolumn{1}{c}{per element} &
  \multicolumn{1}{c}{[\kms]} &
  \multicolumn{1}{c}{[\kms]} &
  \multicolumn{1}{c}{[\kms]} &
  \multicolumn{1}{c}{} & 
  \multicolumn{1}{c}{[arcsec]} &
  \multicolumn{1}{c}{[kpc]}\\
\hline
\multicolumn{9}{c}{Highest SNRs}\\
\hline
J1203+4641 & 0.0723 & 55.6 & $267\pm6$ & $277\pm14$ &       $273\pm14$ & $11.26\pm0.05$ & $1.89\pm0.01$ & $2.69\pm0.02$\\
J2318+1507 & 0.1704 & 49.5 & $226\pm6$ & $226\pm11$ & $224\pm11$ & $11.47\pm0.01$ & $0.92\pm0.01$ & $2.77\pm0.03$\\
J1708+2723 & 0.1058 & 49.3 & $200\pm5$ & $213\pm11$ & $216\pm11$ & $11.07\pm0.02$ & $1.12\pm0.01$ & $2.25\pm0.02$\\
J1426+1647 & 0.0532 & 48.9 & $210\pm4$ & $212\pm11$ & $213\pm11$ & $10.70\pm0.04$ & $0.76\pm0.01$ & $0.82\pm0.01$\\
J1443+0739 & 0.0835 & 48.8 & $234\pm5$ & $228\pm11$ & $227\pm11$ & $11.18\pm0.03$ & $1.64\pm0.02$ & $2.66\pm0.03$\\
J1318+0114 & 0.0785 & 47.7 & $245\pm5$ & $252\pm13$ & $249\pm12$ & $10.89\pm0.01$ & $0.92\pm0.01$ & $1.41\pm0.01$\\
\hline
\multicolumn{9}{c}{Median SNRs}\\
\hline
J1512+4351 & 0.1300 & 22.5 & $209\pm10$ & $215\pm21$ & $210\pm21$ & $10.85\pm0.04$ & $0.54\pm0.02$ & $1.29\pm0.04$\\
J1540+1453 & 0.1134 & 22.4 & $234\pm10$ & $282\pm28$ & $249\pm25$ & $10.96\pm0.01$ & $0.80\pm0.01$ & $1.70\pm0.02$\\
J0040+0332 & 0.1667 & 22.4 & $272\pm12$ & $283\pm28$ & $301\pm30$ & $11.14\pm0.03$ & $0.86\pm0.02$ & $2.52\pm0.05$\\
J1350+2732 & 0.1400 & 22.4 & $229\pm10$ & $230\pm23$ & $238\pm24$ & $10.96\pm0.03$ & $0.68\pm0.01$ & $1.72\pm0.03$\\
J1138+3355 & 0.2379 & 22.4 & $230\pm12$ & $215\pm21$ & $219\pm22$ & $10.93\pm0.05$ & $0.48\pm0.01$ & $1.87\pm0.05$\\
J1550+5627 & 0.1792 & 22.2 & $297\pm12$ & $330\pm33$ & $319\pm32$ & $11.25\pm0.01$ & $0.80\pm0.01$ & $2.49\pm0.04$\\
\hline
\multicolumn{9}{c}{Lowest SNRs}\\
\hline
J1317+3640 & 0.1832 & 13.5 & $255\pm16$ & $255\pm25$ & $250\pm25$ & $11.31\pm0.06$ & $0.98\pm0.03$ & $3.13\pm0.10$\\
J0003+1607 & 0.1518 & 13.1 & $229\pm15$ & $213\pm21$ & $230\pm23$ & $11.00\pm0.03$ & $0.61\pm0.01$ & $1.68\pm0.02$\\
J1035-0045 & 0.1156 & 12.4 & $210\pm14$ & $218\pm22$ & $209\pm21$ & $10.76\pm0.03$ & $0.62\pm0.01$ & $1.33\pm0.03$\\
J0908+0656 & 0.2213 & 11.5 & $325\pm18$ & $341\pm34$ & $332\pm33$ & $11.45\pm0.06$ & $1.00\pm0.04$ & $3.69\pm0.15$\\
J1108+5006 & 0.1169 & 11.4 & $261\pm16$ & $262\pm26$ & $257\pm26$ & $10.90\pm0.02$ & $0.82\pm0.02$ & $1.79\pm0.04$\\
J1436+1624 & 0.0805 & 9.5 & $209\pm18$ & $207\pm21$ & $210\pm21$ & $10.69\pm0.04$ & $0.79\pm0.01$ & $1.23\pm0.01$\\
\hline
\multicolumn{9}{c}{Highest velocity dispersions from \ppxf }\\
\hline
J1328+2757 & 0.1342 & 29.3 & $499\pm17$ & $550\pm55$ & $534\pm53$ & $11.46\pm0.03$ & $1.40\pm0.02$ & $3.45\pm0.04$\\
J0103+1426 & 0.1889 & 23.5 & $435\pm18$ & $496\pm50$ & $489\pm49$ & $11.48\pm0.02$ & $1.19\pm0.02$ & $3.86\pm0.08$\\
J2223-0012 & 0.2934 & 21.7 & $260\pm18$ & $488\pm49$ & $363\pm36$ & $11.55\pm0.03$ & $0.77\pm0.03$ & $3.50\pm0.14$\\
J1022+0521 & 0.1835 & 16.8 & $431\pm20$ & $449\pm45$ & $430\pm43$ & $11.39\pm0.03$ & $1.17\pm0.03$ & $3.72\pm0.09$\\
J0946+4929 & 0.1676 & 19.4 & $363\pm18$ & $437\pm44$ & $414\pm41$ & $11.26\pm0.02$ & $1.02\pm0.03$ & $3.01\pm0.08$\\
J1418+0807 & 0.1426 & 20.9 & $370\pm15$ & $433\pm43$ & $426\pm43$ & $11.23\pm0.02$ & $0.96\pm0.03$ & $2.50\pm0.07$\\
\hline
\multicolumn{9}{c}{Lowest velocity dispersions from \ppxf}\\
\hline
J1126+0612 & 0.2702 & 18.9 & $212\pm11$ & $194\pm19$ & $197\pm20$ & $11.36\pm0.02$ & $0.78\pm0.03$ & $3.35\pm0.12$\\
J1704+2919 & 0.2079 & 22.2 & $208\pm9$ & $194\pm19$ & $212\pm21$ & $11.35\pm0.04$ & $0.99\pm0.03$ & $3.48\pm0.11$\\
J0956+6017 & 0.1142 & 23.0 & $205\pm9$ & $193\pm19$ & $197\pm20$ & $10.70\pm0.01$ & $0.57\pm0.01$ & $1.23\pm0.02$\\
J1346+2440 & 0.1675 & 21.1 & $209\pm15$ & $191\pm19$ & $177\pm18$ & $11.12\pm0.03$ & $0.87\pm0.03$ & $2.57\pm0.09$\\
J1303+3938 & 0.1198 & 24.9 & $201\pm7$ & $191\pm19$ & $194\pm19$ & $10.94\pm0.01$ & $0.85\pm0.02$ & $1.89\pm0.04$\\
J1523+0113 & 0.2376 & 17.3 & $204\pm11$ & $184\pm18$ & $189\pm19$ & $11.29\pm0.03$ & $0.78\pm0.03$ & $3.03\pm0.11$\\
\hline\end{tabular}
    \caption{Kinematic properties for a representative selection of SDSS UCMGs. For each system, we report the redshift, integrated optical SNR, the velocity dispersion value from SDSS and those measured by \ppxf, as described in Section~\ref{sec:kinem}. We also report the stellar mass and effective radii in both arcsec and kpc, the calculation of which is described in Section~\ref{sec:ucmg}. The kinematic properties of all UCMGs in our sample is available in electronic form from the master catalogue.}
    \label{tab:veldisp}
\end{table*}

\subsection{Stellar population analysis}
\label{sec:stel_pop}
As in \citet[\INSPIRE\ DR1]{Spiniello+21} and \citet[\INSPIRE\ DR3]{Spiniello24}, the stellar population analysis is performed in two separate steps. We first carry out a line-index analysis to infer the SSP-equivalent [Mg/Fe] ratios and then use full spectral fitting to estimate mass-weighted (and light-weighted) stellar ages and metallicities. 
This is mainly motivated by the fact that
inferring \afe\ from full-spectral fitting and measuring the [Mg/Fe] from indices does not seem to be the same \citep{Barbosa21, Pernet24}. Measuring \afe\ from full-fitting requires further tests (\citealt{Vazdekis15, Liu20}) and hence we prefer to follow the steps already used in previous \INSPIRE\ papers. This also ensures that the stellar population results and measurements for the two sets of galaxies can be directly compared.

During both steps, we use the same set of single stellar population (SSP) models. Specifically, as already done for the kinematics (Sec.~\ref{sec:kinem}) and in previous papers, we use the MILES models \citep{Vazdekis15} 
with BaSTI theoretical isochrones\footnote{
\url{http://www.oa-teramo.inaf.it/BASTI}.} \citep{Pietrinferni04, Pietrinferni06}. 
From the MILES website\footnote{\url{http://research.iac.es/proyecto/miles/pages/webtools/tune-ssp-models.php}}, we retrieve models with ages ranging from 0.5 to 14 Gyr in steps of 0.5 Gyr, and 10 different values of [M/H] ranging from $-1.49$ dex to $+0.40$ dex, and with the two publicly available values for the \afe\ abundance: 0.0 (solar) and 0.4 (super-solar). 

The biggest assumption we make here concerns the Initial Mass Function (IMF), which we fix to a bimodal power-low with a low-mass end logarithmic slope of 1.3, reproducing what has been done in \citetalias{Spiniello24}. This is broadly equivalent to a Salpeter IMF in terms of mass-to-light ratio estimates. We however point out that in \citet{Maksymowicz-Maciata24} 
we have proven that relics prefer a bottom-heavier IMF than UCMGs with lower DoR. Unfortunately, however, to properly measure the IMF slope from each spectrum 
would require a higher SNR 
than is found in SDSS spectra. We cannot manually check each spectrum like in \citet{Maksymowicz-Maciata24} due to the far greater number of objects and so we are forced to make the assumption of fixed IMF. In order to minimise the effect of this assumption on our results, we restrict the stellar population analysis to wavelengths bluer than 6500\AA, where the contribution to the light from M-dwarf stars is very small \citep{Worthey+94, Spiniello+14}. The relic confirmation that we present in the results section is not affected by this assumption. However, we stress that 
by fixing the IMF, we bias our results by introducing a systematic offset to our derived values of [M/H] and ages, especially for low-DoR systems, (see Appendix B in \citealt{Maksymowicz-Maciata24}). 

In future works we will leverage the results achieved here and stack systems with similar SFHs to increase the resulting SNR and detect variation in the IMF slope at the population level. 

\subsubsection{SSP-equivalent [Mg/Fe] abundances from line-indices}
\label{sec:index}
The [Mg/Fe] abundance, computed from the stellar absorption lines, can be interpreted as the time-scale efficiency of the star formation episode. In fact, 
if the quenching occurs in a very short period, Type Ia supernovae do not have the time to pollute the interstellar medium with iron \citep[e.g.,][]{Matteucci94, Thomas+05, Gallazzi+06, Gallazzi21} and hence the ratio is larger.  Relics have indeed formed through a very fast star formation burst, and therefore 
they have a high, super-solar [Mg/Fe] ratio
\citep{Yildirim17, Ferre-Mateu+17, Martin-Navarro18,Spiniello+21, Spiniello24}. 

We estimate the [Mg/Fe] via the analysis of line-index strengths from Mg and Fe lines. In particular, we take the Mg$_b$ line (5177\AA) and the average of 2 different iron lines (Fe5270 and Fe5335), with the aim of minimising the dependency on other elemental abundances (e.g. [Ti/Fe]) which might fall in the Fe-indices bandpasses\footnote{We have however tested our results against the average of 8 iron lines (Fe4383, Fe4531, Fe5015, Fe5270, Fe5335, Fe5406, Fe5709, Fe5782) and found that the inferred value of \afe\ does not change by more than our uncertainty of 0.1.}.

To calculate the strengths of these features from the galaxies' and the SSP's spectra, we use an in-house code (Benedetti et al. in prep) which is a Python implementation of the \textsc{PACCE} algorithm \citep{Riffel+11}. 
Unfortunately, the MILES models do not allow us to control [Mg/Fe] directly, but this can be approximated to \afe, which is the parameter we can change in the SSPs. 
The version of the MILES models we employ here is only available at two different \afe\ abundances: 0.0 (solar) and 0.4. 
We therefore linearly interpolate the flux at each wavelength, building five sets of models with a $\Delta$\afe=0.1. We prefer not to extrapolate the models outside the original boundaries. 

Differently from what we did in previous publications, here  
we do not convolve data and models to the same resolution. 
Instead, we account for variations in velocity dispersion and spectral resolution by computing `correction factors' $C(\sigma)$, such that the corrected line-index strength is given by $\mathrm{I_{corr}}=\mathrm{I_{orig}}\times C(\sigma)$ \citep{Davies+93, Kuntschner04}, where the I indicates the index under consideration. 
To compute these `correction factors', we first download MILES models with resolutions covering the entire range of $\sigma_{\star}$, as derived in Section~\ref{sec:kinem}. Specifically, we use here models with a single value for metallicity ($\mathrm{[M/H]}=+0.06$ dex), age (11 Gyr) and (\afe $=0.0$). This is motivated by two pieces of evidence. First, we find that the `correction factors' have minimal dependence on these stellar population parameters, especially for $\sigma_{\star}<500$ \kms. Also, we note that in the Mg$_b$-$\langle$Fe$\rangle$ index--index plot, \afe\ (and hence [Mg/Fe]) varies in an orthogonal direction to [M/H] and age. Hence, this index-index plot makes us able to infer the [Mg/Fe], minimising the uncertainties arising from changes in the other stellar population parameters. 
We then define $C(\sigma)$ for each model through $C(\sigma)=\mathrm{I}(60$ \kms$)/\mathrm{I}(\sigma)$. To extend $C(\sigma)$ to all resolutions within the range $60<\sigma_{\star}<550$ \kms (the upper limit is set by the highest velocity dispersion computed from \ppxf\ in Section \ref{sec:kinem}), we use a cubic spline interpolation, following the receipt outlined in \citet{Davies+93}. We list these `correction factors' in Table~\ref{tab:correct}. 

\begin{figure}
    \centering      \includegraphics[width=\columnwidth]{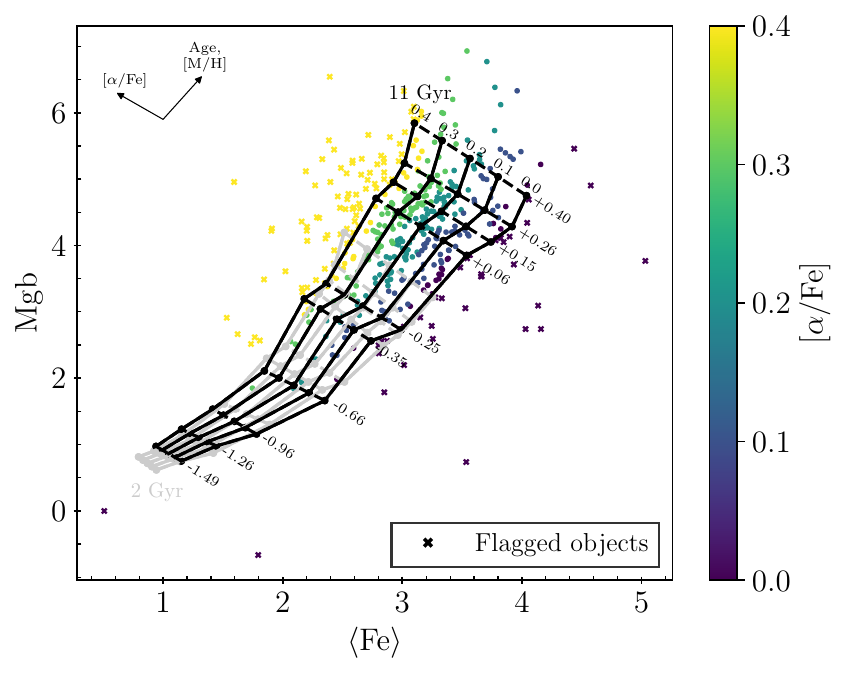}
    \caption{Mg$_b$-$\langle$Fe$\rangle$ index--index plot. The grid shows MILES SSPs with two different ages (2 Gyr in grey and 11 Gyr in black), covering a range of metallicities from -1.49 to +0.40 (solid lines) and a range of \afe\ values between 0.0 and 0.4 (dashed lines). These models have a resolution of 60 \kms. The UCMGs are colour-coded by their derived value of \afe. The direction of variation for age, [M/H], and \afe\ is given by arrows in the top-left corner. Galaxies falling outside the model grid are highlighted with crosses and flagged in the catalogue.}
    \label{fig:alpha}
\end{figure}

\begin{table}
\centering
\begin{tabular}{l|cccc}
    \hline
Correction factor &   Mg$_b$ & Fe5270 & Fe5335 &    $\langle \mathrm{Fe}\rangle$\\
    \hline
$C(200$ \kms) &   1.043 & 1.138 & 1.202 & 1.168  \\
$C(250$ \kms) &   1.090 & 1.207 & 1.339 & 1.269 \\
$C(300$ \kms) &   1.152 & 1.281 & 1.511 & 1.385 \\
$C(400$ \kms) &   1.306 & 1.439 & 1.959 & 1.655 \\
$C(550$ \kms) &   1.596 & 1.740 & 2.850 & 2.153 \\ 
    \hline
    \end{tabular}
    \caption{The `correction factors' used to treat the line-index strengths of the UCMGs, such that they can be directly compared to SSP models with resolution 60 \kms ($C(60$ \kms$=1$)). See Section~\ref{sec:index} for more details.}
    \label{tab:correct}
\end{table}


The overall effect of $C(\sigma)$ is that we can directly compare the UCMGs to SSP models with a resolution of 60 \kms\ (the minimum resolution of the MILES models to within our allowed uncertainty).

The results of the index calculation are shown in Figure~\ref{fig:alpha}. 
The line-index strengths of the models with a resolution of 60 \kms\ are plotted, with models of equal metallicity joined by solid lines and models of equal \afe\ joined by dashed lines. The models with an age of 11 Gyr are in black whilst models with an age of 2 Gyr are shown in grey.
We caution the reader that the models with $\mathrm{[M/H]}=0.40$ dex are outside the `safe ranges'. 
However, we believe we can still use these models in the line-index analysis since [Mg/Fe] varies in an orthogonal direction to [M/H] (see the arrows on the top left corner), which means we can extrapolate beyond the `safe ranges' in the direction of increasing metallicity without affecting the inference on the [Mg/Fe]. 
The UCMGs are plotted over the grids, where their line-index strengths have been corrected to a resolution of 60\kms. 
From the plot, we take as the `true' value for [Mg/Fe] the \afe\ of the closest model to each point on the 11 Gyr grid. Given that there is no prior on age and metallicity and these will slightly shift the model grids, we assign an uncertainty of 0.1 to all the [Mg/Fe] values, which is equal to the steps between the models. 
These ``SSP-equivalent" [Mg/Fe] estimates will be used in the next section to select SSP models with a given \afe\ as input in the full-spectral fitting. There are a number of points falling outside the model grid. The majority of them (80) would be consistent with a larger \afe, making them perfect relic candidates. A negative value of \afe\ (for 36 systems) could indicate instead a more extended SFH. However, we caution that these values might also be due to one or more indices being contaminated by the sky or bad pixel lines. We therefore attach a `flag' to all the systems that lie outside the grid and assign an uncertainty of 0.4 (the maximum possible) to the [Mg/Fe] values of these systems. 

\subsubsection{Mass-weighted stellar ages and metallicities from full-spectral fitting}
\label{sec:stel_pop_ppxf}
Mass-weighted stellar ages and metallicities\footnote{Light-weighted quantities are provided in the master catalogue.} are obtained using \ppxf, which performs a full-spectral fit on the log-rebinned and rest-framed 1D SDSS spectra to which a multiplicative polynomial of degree 8 is applied to correct the shape of the continuum\footnote{In Appendix A of the \citetalias{Spiniello20_Pilot} we performed a test to assess how the stellar population results change when varying the degree of the polynomial. We refer the reader to that paper for a more detailed information.}. For the stellar population fit, the wavelength range is restricted to [3600-6500]\AA\ since it is the cleanest spectral region (as discussed in Section~\ref{sec:kinem}) and the one where varying the low-mass end of the IMF would give a minimal contribution. 
Additionally, we only use models with ages up to the age of the Universe at the redshift of each galaxy, since it would not make physical sense to use models corresponding to older ages.  
Finally, we limit ourselves to models in the `safe range', i.e. metallicities up to $\text{[M/H]}=0.26$ dex. This is different from what we did in previous \INSPIRE\ papers but it is the most conservative choice. In fact, models with 
higher metallicities are not considered `safe' for full-spectral fitting due to the lack of stars in the solar neighbourhood with metallicity higher than $\text{[M/H]}\sim0.2$ dex, upon which the MILES models are based \citep{Vazdekis+12,Vazdekis15}. Nevertheless, we note that the addition of the model with $\text{[M/H]}=0.4$ dex does not heavily affect the age estimates and hence the reconstructed SFHs. We discuss how this influences our results in Appendix~\ref{app:testing}. 

\begin{figure*}
    \centering      \includegraphics[width=\textwidth]{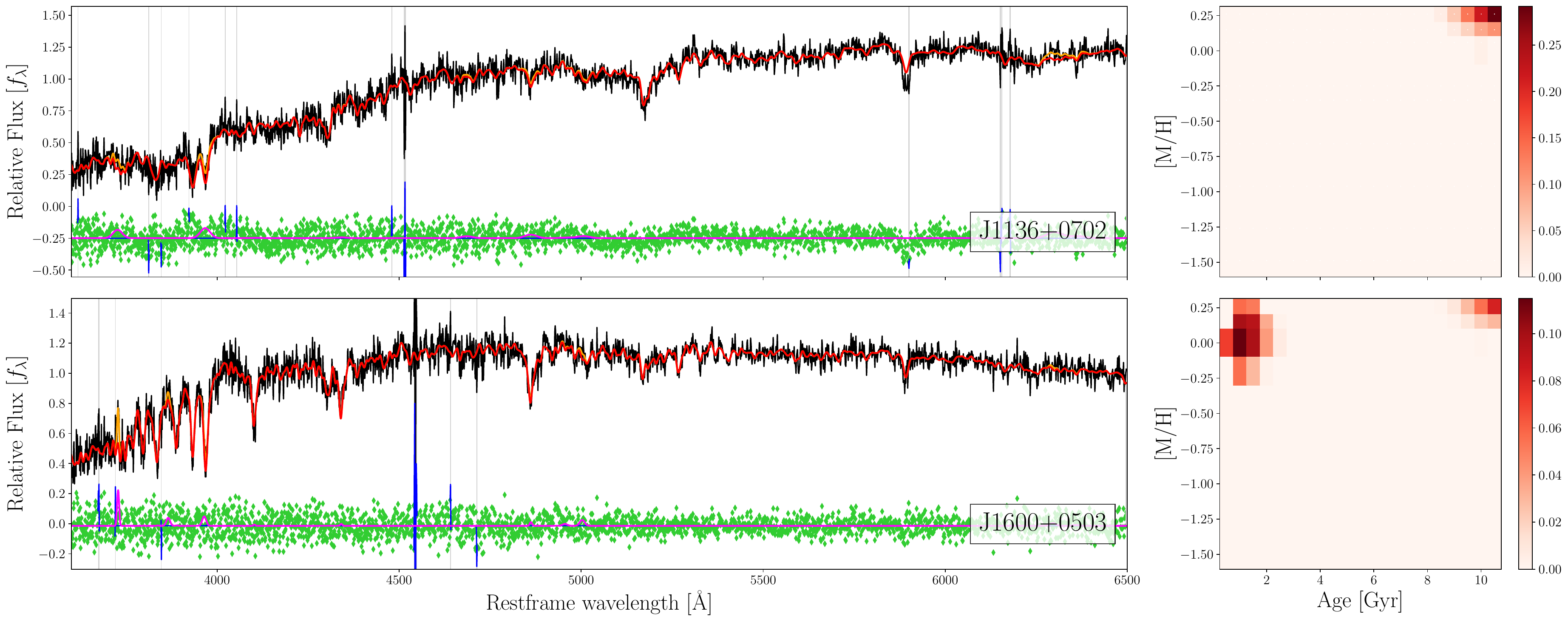}
    \caption{\ppxf\ fits for two galaxies with very different SFHs. \textit{Left panel:} the fits with the galaxy spectra in black, the SSP best-fitting template in red, residuals in green,  pixels masked out from the fit in blue, and gas lines and fit in orange and purple, respectively.  
    \textit{Right panel:} the mass-weighted age–metallicity density maps for the non-bootstrapped (and lightly regularised) fits.}
    \label{fig:stelpop}
\end{figure*}

Similarly to \citetalias{Spiniello24}, for each UCMG, we perform four different \ppxf\ runs as described below. In all cases, we work in the 2D space where only age and metallicity are free to vary during the fits. Furthermore, in all the \ppxf\ fits, we also fit gas lines (purple lines in Figure~\ref{fig:stelpop}), using the standard list given by \ppxf\ as these might influence the age and metallicity estimates, especially when the emissions are superimposed on stellar absorptions.  
However, due to the large number of UCMGs, we cannot proceed in the same way as previous \INSPIRE\ papers where we used the regularisation method, repeating the fit multiple times for each galaxy until finding the maximum regularising factor allowed by the data (see \citetalias{Spiniello+21} and \citetalias{Spiniello24} for more information). Hence, we use a slightly different approach, as described below. First, we perform 
a lightly regularised fit with a regularisation of 10 for all systems, using models with \afe\ equal to the [Mg/Fe] SSP-equivalent values found from the index-index analysis (see Section~\ref{sec:index}). Then, since the inferred \afe\ have an uncertainty of 0.1, we run \ppxf\ again, once with \afe=[Mg/Fe]+0.1 and once with \afe=[Mg/Fe]-0.1. For 'flagged objects', we instead perform these fits using models with \afe=0.0 and \afe=0.4, hence deriving the most extreme SFHs for these objects. 

At this point, we perform an unregularised fit with the `true' \afe\ and use a bootstrapping method ('wild bootstrapping', described in \citealt{DavidsonFlachaire08}) to obtain a distribution of the weights assigned to each SSP template. This is similar to, although much quicker than, the regularisation process which smooths the stellar population solution as much as the data allows. The main difference is that when performing regularised fits, one must manually (or iteratively) try different regularisation values, each time rescaling the noise level, until reaching the MAX\_REGUL for which the $\chi^2$ increases by $\sqrt{2 \times N_{\rm pixel}}$ (where $N_{\rm pixel}$ is the number of pixel used in the fit). The MAX\_REGUL changes system by system, which would have make this procedure very time consuming with $\sim400$ galaxies. We follow the procedure recommended by the author of \ppxf\ where we use bootstrapping of the residuals, while repeating the \ppxf\ fits 9 times to obtain averages for the stellar ages and metallicities \citep{Cappellari17, Cappellari23}. Note that we run the fits where \afe\ is varied from the `true' value without bootstrapping. We also tested different numbers of iterations for the bootstrapping routine (up to 20), finding negligible differences on the distribution of the weights.
We also test that the results obtained with the bootstrapping procedure are compatible with the ones obtained with the regularisation, on ten randomly selected spectra. 

We finally calculate mean values and standard deviations for the mass-weighted stellar ages and metallicities from all the 
fits performed for each system (the three non-bootstrapped runs with varying values of \afe\ and the bootstrapped run). These are listed, along with the [Mg/Fe] ratios derived in Section~\ref{sec:index}, 
for a selection of UCMGs in Table~\ref{tab:stelpop}. The same quantities for the entire sample are available in the catalogue released online as part of this publication. 
Figure~\ref{fig:stelpop} shows two 
examples of (non-bootstrapped) \ppxf\ fits for objects with very different SFHs. The galaxy spectra are plotted on the left and the weights of the age and metallicities are plotted on the right. The upper galaxy is much older and had all of its stellar mass formed in one burst (i.e. it is a relic). On the other hand, the galaxy on the bottom panel is much younger and has a much more extended SFH.

\begin{table}
\centering
\begin{tabular}{lcrr}
\hline
  \multicolumn{1}{c}{ID} &
  \multicolumn{1}{c}{[Mg/Fe]} &
  \multicolumn{1}{c}{$t_{\mathrm{mean}}$} &
  \multicolumn{1}{c}{[M/H]$_{\mathrm{mean}}$} \\
  \multicolumn{1}{c}{SDSS} &
  \multicolumn{1}{c}{(dex)} &
  \multicolumn{1}{c}{(Gyr)} &
  \multicolumn{1}{c}{(dex)} \\
\hline
\multicolumn{4}{c}{Highest stellar ages}\\
\hline
J1616+2809 & 0.1 & $12.4\pm0.1$ & $0.224\pm0.013$\\
J1358+1327 & 0.3 & $12.4\pm0.2$ & $0.235\pm0.004$\\
J0750+2259 & 0.1 & $12.3\pm0.3$ & $0.235\pm0.007$\\
J1433+5315 & 0.2 & $12.2\pm0.4$ & $0.235\pm0.001$\\
J1426+1647 & 0.2 & $12.2\pm0.3$ & $0.236\pm0.002$\\
J1510+0546 & 0.3 & $12.1\pm0.1$ & $0.240\pm0.004$\\
\hline
\multicolumn{4}{c}{Median stellar ages}\\
\hline
J2146-0821 & 0.2 & $10.4\pm0.5$ & $0.209\pm0.016$\\
J0824+1038 & 0.3 & $10.4\pm0.6$ & $0.173\pm0.014$\\
J1107+6506 & 0.1 & $10.4\pm0.5$ & $0.216\pm0.010$\\
J1427+2107 & 0.4 & $10.4\pm1.1$ & $0.162\pm0.034$\\
J1520+2035 & 0.2 & $10.4\pm0.4$ & $0.210\pm0.011$\\
J1338+0541 & 0.3 & $10.4\pm0.1$ & $0.224\pm.004$\\
\hline
\multicolumn{4}{c}{Lowest stellar ages}\\
\hline
J1126+0612 & 0.0 & $3.5\pm2.0$ & $0.114\pm0.078$\\
J0905+3657 & 0.1 & $3.3\pm0.8$ & $0.141\pm0.034$\\
J2305-1033 & 0.1 & $3.2\pm1.2$ & $0.145\pm0.050$\\
J1602+0245 & 0.2 & $3.1\pm2.1$ & $0.115\pm0.071$\\
J1452+0253 & 0.0 & $2.9\pm2.4$ & $0.145\pm0.088$\\
J1600+0503 & 0.0 & $2.3\pm1.5$ & $0.061\pm0.046$\\
\hline
\multicolumn{4}{c}{Highest metallicities}\\
\hline
J1620+1722 & 0.2 & $11.6\pm0.1$ & $0.240\pm0.007$\\
J1710+3941 & 0.2 & $11.8\pm0.4$ & $0.240\pm0.002$\\
J2130+0307 & 0.2 & $12.1\pm0.1$ & $0.240\pm0.003$\\
J1508+6049 & 0.3 & $12.0\pm0.2$ & $0.240\pm0.007$\\
J1732+3102 & 0.3 & $11.5\pm0.2$ & $0.239\pm0.002$\\
\hline
\multicolumn{4}{c}{Median metallicities}\\
\hline
J1122+2017 & 0.2 & $10.4\pm0.6$ & $0.204\pm0.011$\\
J0752+3514 & 0.1 & $9.0\pm1.6$ & $0.203\pm0.010$\\
J0736+4336 & 0.0 & $9.8\pm0.6$ & $0.203\pm0.016$\\
J1540+3454 & 0.4 & $9.9\pm0.2$ & $0.203\pm0.015$\\
J1206+6207 & 0.1 & $7.2\pm1.6$ & $0.202\pm0.026$\\
J1505+3007 & 0.2 & $8.5\pm2.5$ & $0.202\pm0.033$\\
\hline
\multicolumn{4}{c}{Lowest metallicities}\\
\hline
J1227-0304 & 0.3 & $9.0\pm1.2$ & $-0.058\pm0.188$\\
J1511+3020 & 0.4 & $5.4\pm1.7$ & $-0.100\pm0.074$\\
J0901+2242 & 0.0 & $9.1\pm0.5$ & $-0.125\pm0.164$\\
J2223-0012 & 0.3 & $6.2\pm2.1$ & $-0.211\pm0.141$\\
J1357+1553 & 0.2 & $6.7\pm1.0$ & $-0.231\pm0.238$\\
J0815+0635 & 0.0 & $5.4\pm1.3$ & $-0.320\pm0.129$\\
\hline\end{tabular}
    \caption{Stellar population results for a representative selection of systems. For each system, we also report the SSP-equivalent [Mg/Fe]. We estimate the stellar age and metallicity by taking the mean of all the fits for each system (see Section~\ref{sec:stel_pop_ppxf}). Note that J1510+0546 is both amongst the oldest and most metal rich systems, and so is only included once on the table for conciseness.}
    \label{tab:stelpop}
\end{table}

\section{Results}
\label{sec:results}
In this section we use the stellar population parameters computed from the analysis described in Section~\ref{sec:analysis} to reconstruct the star formation histories of the UCMGs, as well as their stellar metallicity evolutions in cosmic time. This allows us to divide them into different groupings, computing their \textit{Degree of Relicness} and investigating how it relates to their stellar population properties. We will discuss how the assumptions we made influence the DoR in the dedicated Appendix~\ref{app:testing}.

\subsection{Time evolution of mass and metallicity}
\label{sec:sfh}
From the weights of the SSP models \ppxf\ uses in the fit, we can infer the cosmic evolution of the stellar mass (i.e. the fraction of stellar mass assembled from the Big Bang to the present-day) and of the stellar metallicity. 

To calculate the cumulative stellar mass assembled in cosmic time (i.e. the SFH), we follow the receipt from previous \INSPIRE\ papers. We start from the density maps, like the ones shown in the right panels of Figure~\ref{fig:stelpop}. We flip the age axis and sum over all metallicity values, computing in this way the fraction of mass assembled at each age bin since the Big Bang to the redshift of the galaxy. 

To calculate the metallicity of a galaxy as a function of cosmic time (the metallicity evolution history or MEH), we follow instead \citealt{Bevacqua+24} (specifically Eq.~5). For each cosmic time, we calculate the corresponding metallicity by computing the weighted mean of the metallicities of the subset of SSP models whose ages are older than that cosmic time. 
To build up the MEH curves, at the earliest cosmic time, we only consider the models with the oldest ages, whereas for `today', i.e. at the redshift of the galaxy, we consider all the models  
so that the metallicity matches with the 
metallicity computed in Section~\ref{sec:stel_pop_ppxf}.

We note that this method of computing the cumulative mass and the metallicity at a given cosmic time assumes that stars form in a series of bursts rather than over an extended period of time (i.e. we use SSPs rather than models with time-declining star formation). 

\begin{figure*}
    \centering
    \includegraphics[width=\textwidth]{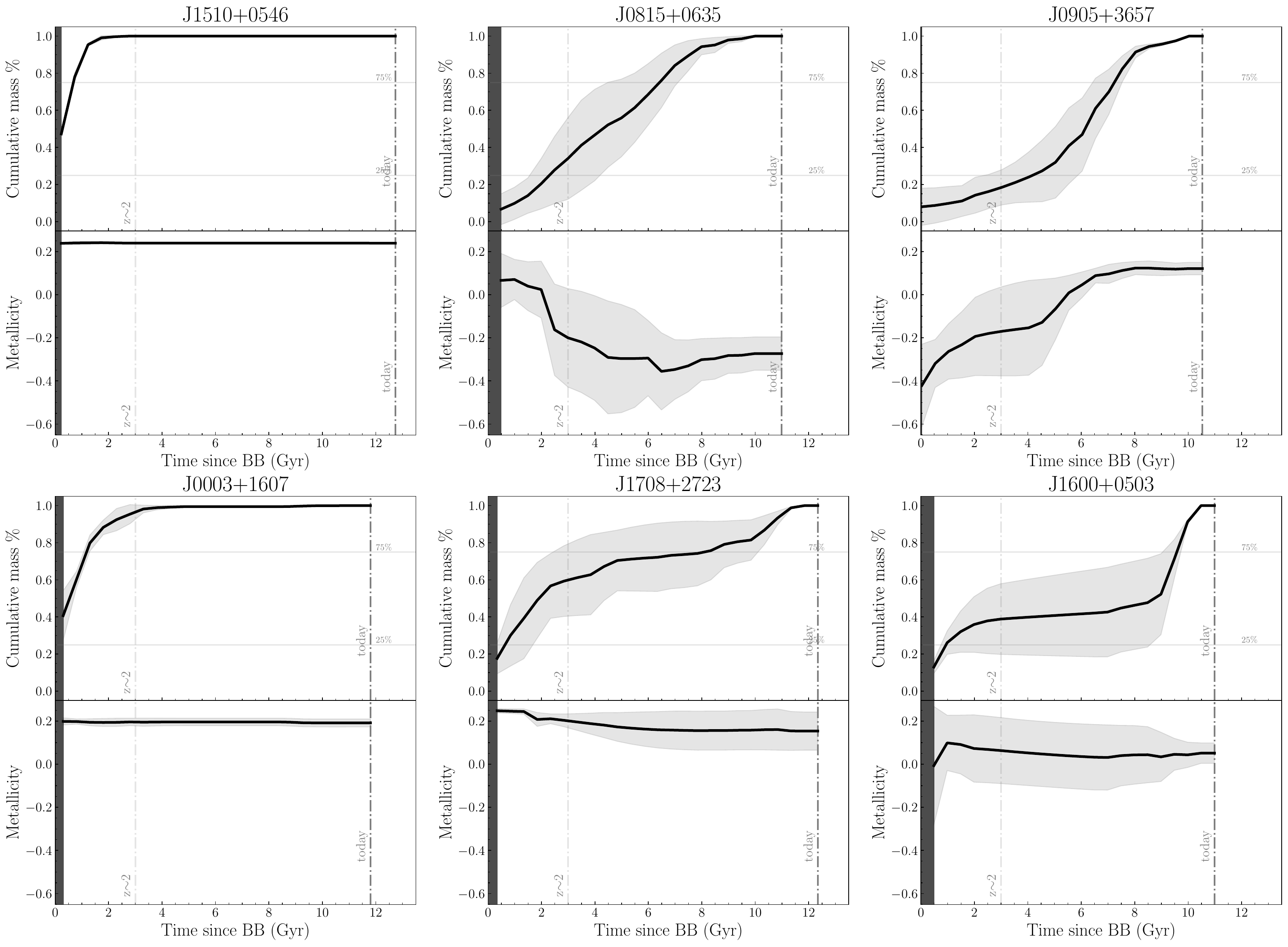}
    \caption{Stellar mass assembly and metallicity evolution in cosmic time, from the BB to the redshift of the galaxy, for a selection of six galaxies. In the top panel of each row, the black line shows the mean stellar mass at each cosmic time of the four fits (as described in Section~\ref{sec:stel_pop_ppxf}). In the bottom panel of each row, the black line show the mean metallicity at each cosmic time of the four fits (also as described in Section~\ref{sec:stel_pop_ppxf}). In both rows, the shaded region is one standard deviation around the mean for each cosmic time. The ID of each galaxy is reported above each galaxy. }
    \label{fig:relics_sfh_meh}
\end{figure*}

A selection of SFHs (top) and MEHs (bottom) are shown in Figure~\ref{fig:relics_sfh_meh} for objects that are representative of the entire sample. 
For each galaxy, whose ID is given in the title, the black line shows the mean quantity (stellar mass or metallicity) at each cosmic time and the shaded region represents one standard deviation around it. Vertical grey lines show the end of the first phase of the formation scenario ($z\sim2$, \citealt{Zolotov15}) and the age of the Universe at the redshift of the system (today). In the top panels, the horizontal lines highlight the 25\% and 75\% thresholds on the stellar mass to guide the eyes. 
From left to right, the objects vary from UCMGs with very peaked and quick SFHs that occurred very early on in cosmic time (i.e. extreme relics), to UCMGs with much more extended SFHs that completed their stellar mass assembly only very recently.

We note that the SSPs' binning is 0.5 Gyr and that we use, for each galaxy, only models with ages up to the age of the Universe at the redshift of the systems. 
We therefore cannot deduce anything about the SFH or the MEH of the galaxies for the cosmic time between the Big Bang and the formation of the oldest SSP models and so this period is masked out in Figure~\ref{fig:relics_sfh_meh}. The exact region which is masked out varies depending on the age of the Universe at the redshift of the galaxy. 


The MEHs show a wide range of behaviours, 
even for objects with similar SFHs. This is particularly true for UCMGs with more extended SFHs (the two right-most columns in Figure~\ref{fig:relics_sfh_meh}), where we observe varying behaviours, resulting in objects spanning a wide range of metallicities across cosmic time. There are also cases where the metallicity stays roughly constant despite the mass increasing and vice versa, where the metallicity changes drastically despite the mass staying roughly constant. In contrast, relics and especially the most extreme ones, tend to have metallicities that stay fairly constant and super-solar at all cosmic times. As we will discuss later, this reflects the fact that relics are generally metal-richer than younger UCMGs. Spatially resolved data will be necessary to further investigate the metallicity, its relation with SFHs, and its evolution with cosmic time.

\subsection{DoR and relic confirmation}
\label{sec:dor}
Following the same approach as in \citetalias{Spiniello24}, we compute for each UCMG the \textit{Degree of Relicness} (DoR) as follows:

\begin{equation}
    \text{DoR} = \left[f_{M^{\star}_{t\text{BB}=3}}+\frac{0.5~\text{Gyr}}{t_{75}}+\frac{0.7~\text{Gyr}+(t_{\rm Uni}-t_{\rm fin})}{t_{\rm Uni}}\right]\times\frac{1}{3},
\end{equation}

where \Mfrac is the fraction of stellar mass at $z\sim2$, $t_{75}$ is the cosmic time at which 75 per cent of the stellar mass was in place, $t_{\rm fin}$ is the final assembly time (100 per cent of the stellar mass in place), and $t_{\rm Uni}$ is the age of the Universe at the redshift of the objects. The values 0.5~Gyr and 0.7~Gyr are chosen such that the DoR ranges between 0 and 1 (as explained in \citetalias{Spiniello24}). A higher DoR indicates an object with a very peaked SFH that formed most of its stellar mass by $z\sim2$ (i.e. a relic), whereas a lower DoR indicates an object with a much more extended SFH with star formation continuing until recently (or still ongoing).
We caution the reader that the DoR is an arbitrarily computed, dimensionless number which is useful to weigh the fraction of very old, `pristine' stars and hence identify the most extreme relics, but that depends on the choices and assumptions made during the fit, as shown in Appendix~\ref{app:testing}. 

We first determine estimates of the parameters \Mfrac, $t_{75}$, and $t_{\rm fin}$ for the four \ppxf\ fits performed for each object (as described in Section~\ref{sec:stel_pop_ppxf}). We then choose the estimates which give the most conservative (lowest) estimate of the DoR, i.e. we use the minimum value of \Mfrac and the maximum values of $t_{75}$ and $t_{\rm fin}$\footnote{The $t_{\rm fin}$ is defined as the cosmic time at which 99.8\% of the stellar mass is formed. This (conservative) choice is motivated by the findings of \citet{Salvador-Rusinol22} who found that even in the most extreme local relic, NGC1277, a sub-percentage of younger stars are found.}, to reproduce exactly what was done in previous papers. We however caution the readers that these are arbitrary choices that have an influence on the final DoR distribution, as we highlight in Appendix~\ref{app:testing}. Nevertheless we stress that, although the numerical values and normalisation of the DoR is dependent on the assumptions, this quantity in all cases is able to describe the variety of SFHs that, in turn, corresponds to a wide range of systems, from the most extreme relics to relatively younger UCMGs that undergone a more complex and time-extended formation history. 

\begin{figure}
    \centering  \includegraphics[width=\columnwidth]{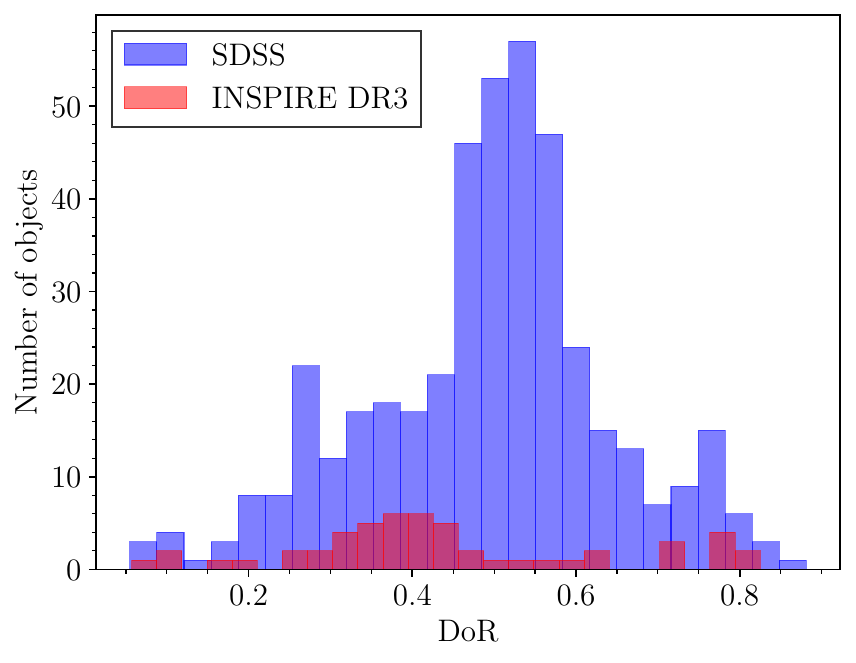}
    \caption{A comparison of the distribution of DoR values for the objects in this paper and the objects found previously in \citetalias{Spiniello24}.}
    \label{fig:dor}
\end{figure}

The distribution of the resulting DoRs for the 430 UCMGs is shown in Figure~\ref{fig:dor} (blue), where we also compare it 
with that of the \INSPIRE\ galaxies (red). We note that there are a number of differences between the approach we follow here and that followed in previous \INSPIRE\ publications. 
First, we use a different selection criterion to identify UCMGs which is based on density, rather than cuts at fixed stellar mass and radius. We do however limit ourselves to objects that can be considered as UCMGs according to the \INSPIRE\ definition, by correcting the SDSS sizes to KiDS resolution and excluding objects with \Reff$_{\rm KiDS}>2$ kpc and by excluding objects with \Mstar$<6\times10^{10}$\Msun. Furthermore, here we decided to exclude the metal-richest SSPs while we included them in previous papers. We investigate the effect of this difference in Appendix~\ref{app:testing} (Figures~\ref{fig:z_scatter} and ~\ref{fig:dor_z}).

We observe a peak in the distribution at $\text{DoR}\sim0.5$, which is slightly higher than in \citetalias{Spiniello24} where the peak is at $\text{DoR}\sim0.4$. However, both distributions cover a similar range of DoR values: $0.05<\text{DoR}<0.9$, reflecting the great variety of SFHs covered by both samples. 


\begin{figure}
    \centering    \includegraphics[width=\columnwidth]{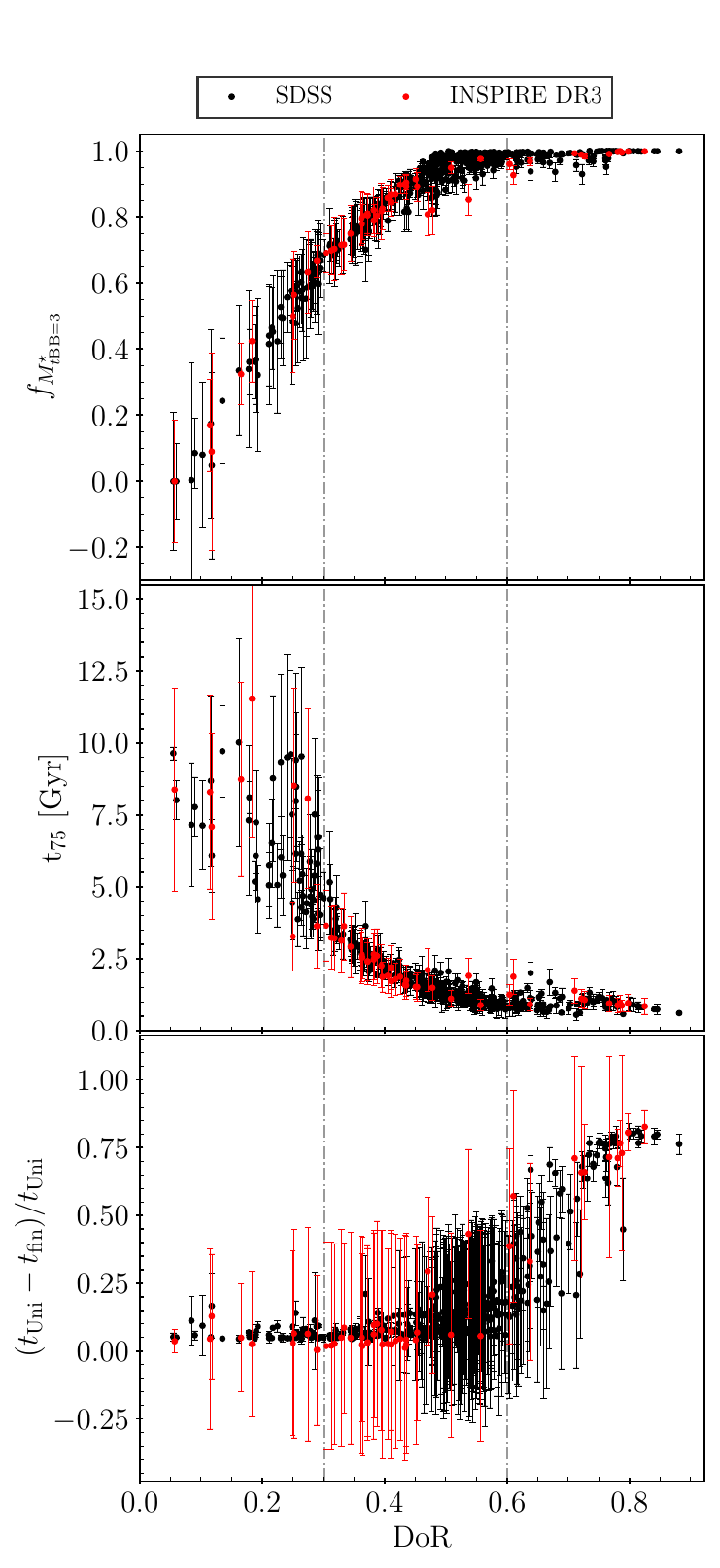}
    \caption{The DoR plotted against the fraction of stellar mass formed by $z\sim2$ (top), the cosmic time at which 75 per cent of the stellar mass was in place (middle), and the time of final assembly (bottom). We compare these values computed for objects in SDSS (black) with those computed for objects in \citetalias{Spiniello24} (red). We also draw vertical lines at $\text{DoR}=0.3$ and $\text{DoR}=0.6$ to indicate changes in behaviour of each quantity, as described in Section~\ref{sec:dor}.}
    \label{fig:dor_prop}
\end{figure}

Figure~\ref{fig:dor_prop} shows the DoR plotted against \Mfrac (top), $t_{75}$ (middle), and $(t_{\rm Uni}-t_{\rm fin})/t_{\rm Uni}$ (bottom) for both the objects in this paper and those in \citetalias{Spiniello24}.  
We remind the readers that the latter quantity is used to take into account the fact that the systems cover a range of redshifts ($0.01<z<0.3$). Hence renormalising by the age of the Universe at each redshift allows us to directly compare the final assembly times of the different objects. 
However, this is also the quantity with the largest uncertainty, which is due to the fact that we set a very restrictive threshold ($>99.8\%$ of the stellar mass in place). We investigate the impact of a slightly different choice for $t_{\rm fin}$ in Appendix~\ref{app:testing}. 
Errors on the plotted quantities are determined by taking the standard deviation of the four estimates of each parameter. 
The 430 UCMGs found in this paper follow a very similar distribution to the objects in \citetalias{Spiniello24}. 

In previous \INSPIRE\ papers, we define relics as objects that had formed 75\% or more than their stellar masses by $z=2$. Using this operative and \textit{'ad-hoc'} threshold, the lowest DoR at which we found a relic was 0.34. Following the same approach, hence considering only objects that formed 75 or more percent of their mass during the first phase of the formation scenario, we would retrieve 359/414 UMCGs and, quite nicely, the lowest DoR would be fully consistent with the previous one (0.346).  
However, thanks to the much larger number statistics, we can now better distinguish the different behaviours for different DoR ranges, hence moving away from an arbitrary mass threshold. We distinguish three different regions, as indicated by the vertical dashed lines in Figure~\ref{fig:dor_prop}. In particular we observe that 
\begin{itemize}
    \item $\text{DoR}\lesssim0.3$. In this range, \Mfrac linearly increases with the DoR but is always $<75$\%, $t_{75}\sim8$ Gyr (i.e. a constant) with a large scatter, and $t_{\rm fin}\sim t_{\rm Uni}$. These UCMGs had very extended SFHs and cannot be considered relics of the ancient Universe. We find 56 of these objects.
    \item $0.3\lesssim\text{DoR}\lesssim0.6$. In this case, \Mfrac keeps increasing with the DoR, although at a slower rate. In many cases, it reaches $\sim100$\% around DoR$\sim0.5$, where a bending in \Mfrac is visible. $t_{\rm fin}$ is still $\sim t_{\rm Uni}$, but now $t_{75}$ starts to decrease with the DoR and the overall scatter is much smaller. These objects, although having formed the great majority of their stellar mass early on in cosmic time, still have a non-negligible fraction of stars that formed later on and/or through time-extended processes. There are 293 of these objects in our catalogue. 
    \item $\text{DoR}\gtrsim0.6$. For these objects, \Mfrac$\sim100\%$, $t_{75}\sim1~\text{Gyr}$, and $(t_{\rm Uni}-t_{\rm fin})/t_{\rm Uni}$ increases linearly with the DoR. Furthermore, above this threshold, the scatter in all three quantities becomes very small and the entire totality of the stellar mass budget is dominated by stars almost as old as the Universe. We find 81
 of these extreme relics.  
\end{itemize}

We note that the second threshold was slightly higher with the original sample of 52 UCMGs (extreme relics were defined to have DoR$\ge0.7$). Now, having increased the sample by almost a factor of 10, we reconsider the groupings based on the new distributions. 

\citet{Grebol2023} analysed a sample of 37 compact galaxies with MaNGA \citep{Bundy15} data, selected to bridge the stellar mass gap between compact elliptical galaxies ($8\le
\log$[\Mstar$/M_{\odot}]\le10$) and UCMGs ($10\le
\log$[\Mstar$/M_{\odot}]$). Using a machine learning based clustering algorithm ($k$-means), they classified the compact galaxies into three groups, according to their stellar properties. In particular, they used the cosmic time at which 90 and 50 percent of the stars were formed, the stellar mass, metallicity and $\Sigma_{1.5}$ value for the classification. Interestingly, the classification we obtained is incredibly similar to the one they derived. 
They found that 76\% of their sample was made of old and metal rich galaxies with extremely steep SFHs. The remainder of the sample is almost equally split between intermediate-age galaxies with a wide range of metallicites (13\%) and young galaxies, showing multiple episodes of star formations (11\%). Both these groups are characterised by slightly $\alpha$-enhanced stars, with lower [$\alpha$/Fe] ratios than the first group.



\begin{figure*}
    \centering    \includegraphics[width=\textwidth]{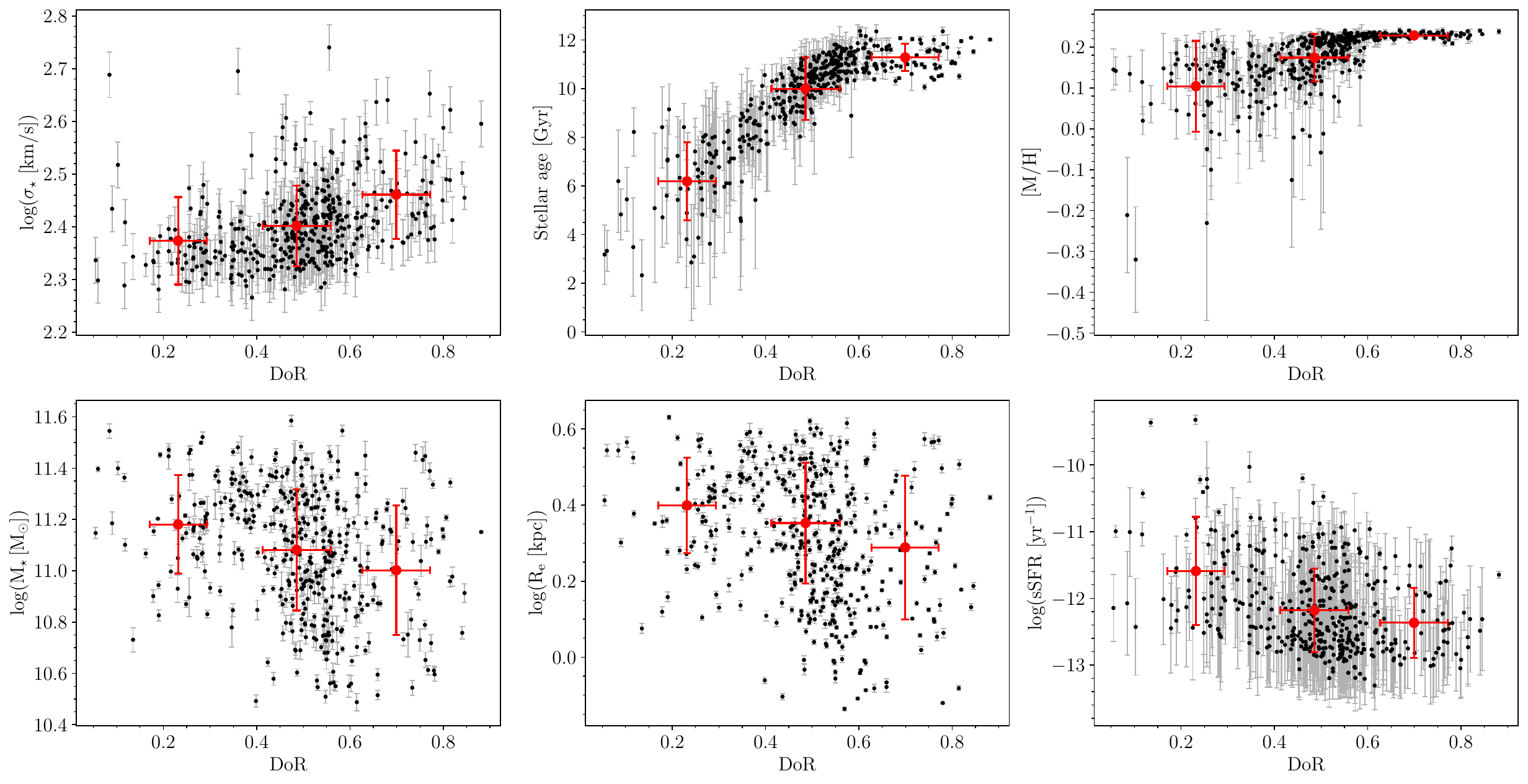}
    \caption{The relationship between the DoR and (from top left to bottom right) stellar velocity dispersion, age, metallicity, stellar mass, effective radii (as defined in Section~\ref{sec:ucmg}), and specific SFRs. The objects have also been grouped by DoR, following the groupings outlined in Section~\ref{sec:dor}. For each group, the mean and standard deviation have been plotted in the panels as red circles with errorbars.}
    \label{fig:dor_corr}
\end{figure*}

\begin{figure}
    \centering
    \includegraphics[width=\columnwidth]{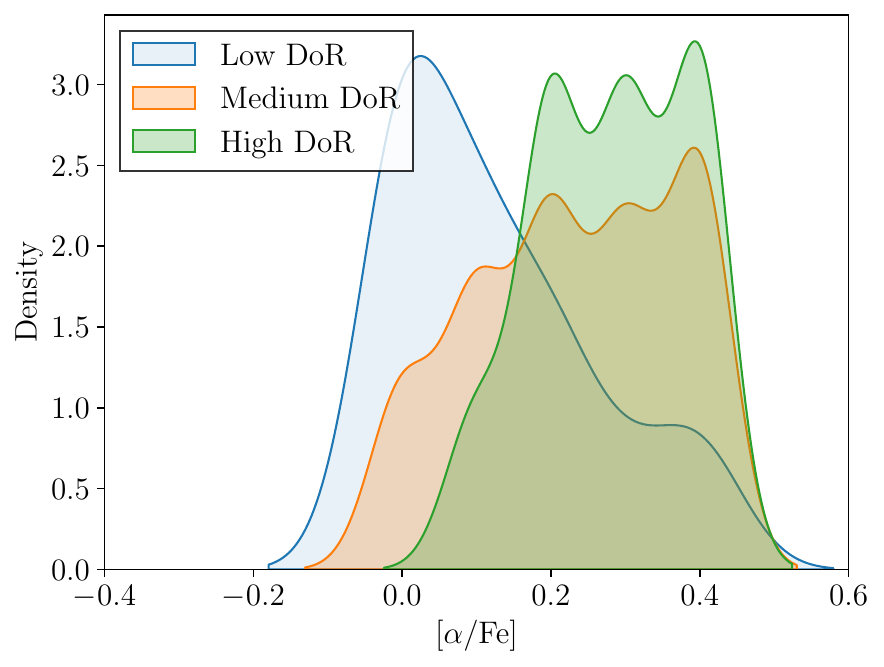}
    \caption{The distribution of \afe\ for the 430 UCMGS. These have been split into three groupings based on DoR, as described in Section~\ref{sec:dor}. The `Low DoR' groups contains objects with $\rm DoR\lesssim0.3$, the `High DoR group contains objects with $\rm DoR\gtrsim0.6$, and the `Medium DoR' group contains everything in between ($0.3\lesssim\text{DoR}\lesssim0.6$). This plot is produced by smoothing the distribution with a Gaussian kernel to improve visibility.}
    \label{fig:alpha_hist}
\end{figure}

In Figure~\ref{fig:dor_corr}, we show the relation between the DoR and a selection of other quantities. Although with a large scatter, we confirm the previous findings that UCMGs with higher DoR are generally metal richer, are older (by construction), and have lower sSFRs. We also find that there is a negligible correlation between DoR and \Reff.
Furthermore, a weak indication that they also have larger stellar velocity dispersions is hinted at in the plot. This is however only evident for $\text{DoR}>0.6$, which is fully consistent with what we found in \citetalias{Spiniello24} (specifically in Fig.~9). 
Interestingly, around the same value of $\text{DoR}>0.6$, the scatter in [M/H] sharply decreases and all the UCMGs are consistent with a super-solar metallicity, reaching the maximum value allowed by the MILES models ($\text{[M/H]}=0.26$ dex). The distribution of \afe\ for the UCMGs is shown in Figure~\ref{fig:alpha_hist}, where we split the objects into the three groupings based on DoR described earlier in this section\footnote{The peaks in the distribution are artificially created by the \afe\ sampling of 0.1.}. We find that, in general, objects belonging to the group with the lowest DoRs have lower \afe\ values than the other two groups. Additionally, objects with the highest DoR values ($\rm DoR>0.6$) have the highest \afe\ values on average, confirming previous findings that UCMGs with higher DoR have slightly higher \afe\ ratios. This suggests that a high [Mg/Fe] value (from which \afe\ is inferred) is an additional good criterion from which to search for relics. 

\begin{figure}
    \centering    \includegraphics[width=\columnwidth]{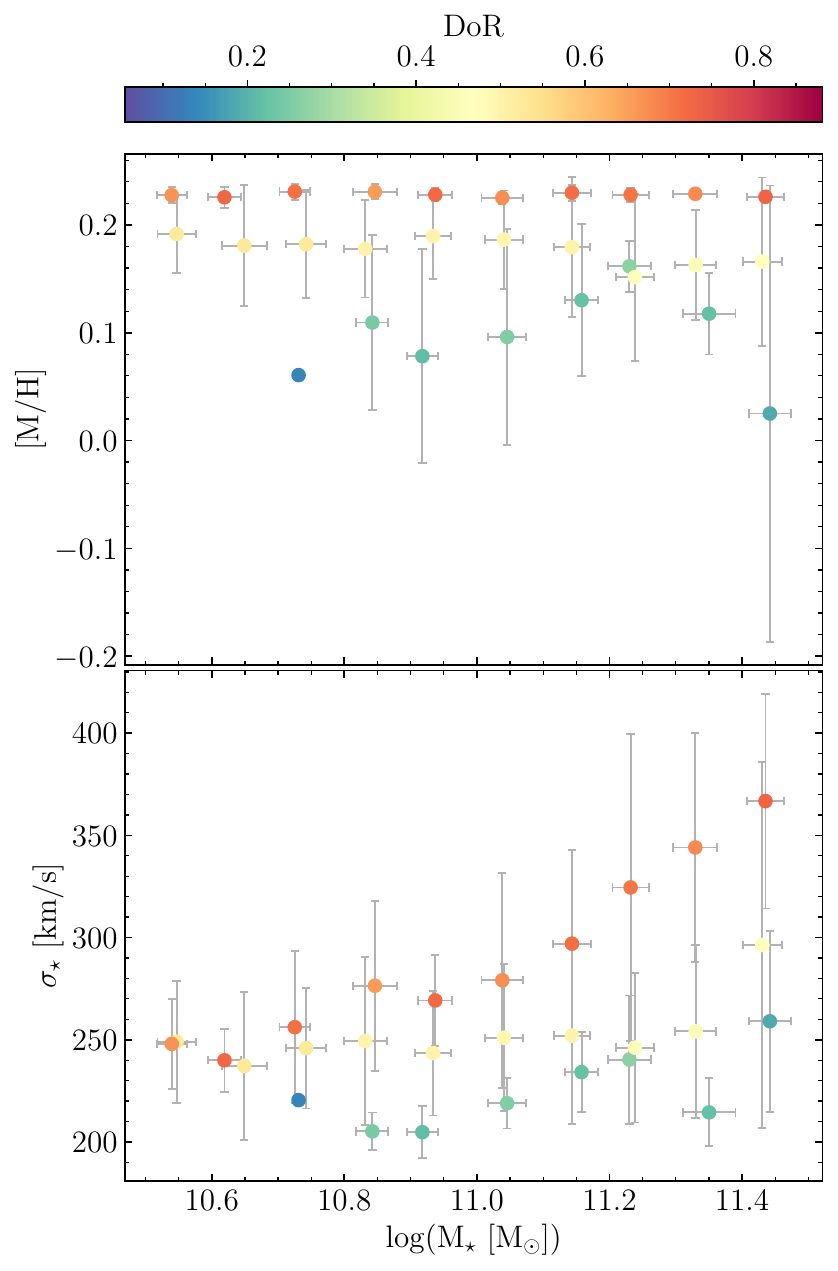}
    \caption{Plots of stellar mass against metallicity (top panel) and velocity dispersion (bottom panel). As described in Section~\ref{sec:dor}, the UCMGs are divided into groups of similar DoR (based on the discussion in that section) and stellar mass. The mean and standard deviation of each group are then plotted on the above figure.}
    \label{fig:dor_mass}
\end{figure}

Since the relation between DoR and metallicity is tighter than that between DoR and $\sigma_{\star}$ or DoR and \afe, we suggest that an even better way to select relics from a large sample of spectroscopically confirmed UCMGs is to select metal-rich objects. To further illustrate this, in Figure~\ref{fig:dor_mass}, we divide the UCMGs into three groups (based on the subdivisions described earlier) and then further divide each group into bins of different mass. For each subdivision we compute the mean and standard deviation of the stellar mass, metallicity (top panel), and velocity dispersion (bottom panel) and plot these points. It is very clear that objects at the top of these distributions tend to have higher DoRs, especially for the metallicity against stellar mass distribution. In this case, the most extreme relics at all mass bins saturate at the maximum metallicity value allowed by the SSP models in their `safe' ranges. 

\section{Conclusions}
\label{sec:conclusions}
This paper is the first of the Extending the INvestigation of Stellar Population In RElics (\EINSPIRE) project, which expands on the original \INSPIRE\ sample with the goal of understanding relics, their formation, time evolutions and environment. In this first paper, we have pushed the redshift boundaries towards the local Universe and extended the original sample of UCMGs with a measured \textit{Degree of Relicness} \citep{Spiniello24} by a factor of $\sim10$, hence bridging the gap with the local Universe. This has been made possible by the much larger sky area covered by the SDSS survey, with respect to KiDS. 
We started by selecting galaxies from SDSS DR18 \citep{Almeida+23_SDSS} at $z<0.4$ with red colours and high stellar velocity dispersion ($\sigma_{\star}>200$ \kms). Furthermore, we used stellar mass estimates from the GSWLC-2 catalogue described in \citet{Salim+18} and estimated the effective radii by combining de Vaucouleurs and exponential radii, directly provided by SDSS. Since the GSWLC-2 only goes up to $z<0.3$, we further reduce the upper redshift boundary of our final sample. We then selected the densest and most compact objects using the criterion of \citet{Baldry21}, first defined in \citet{Barro+13}. We finally assembled a catalogue of 430 spectroscopically confirmed UCMGs, after further cleaning the sample by manually inspecting their stamps and spectra and correcting for the worse spatial resolution of SDSS compared to KiDS. To our knowledge, at the time of writing, this is the largest homogeneously selected catalogue of spectroscopically confirmed UCMGs at $0.01<z<0.3$. The catalogue is publicly available for the entire scientific community and includes coordinates, Petrosian $r$-band magnitudes, stellar masses, specific star formation rates, and effective radii. These quantities are all computed from optical photometric data. In addition, we provide integrated stellar velocity dispersion values as well as stellar population parameters that we estimated from SDSS spectra, as described in Section~\ref{sec:analysis}. Finally, the DoR and the ingredients used to compute it, as well as the stellar population parameters, are listed in the catalogue for all the objects. 

Through a spectroscopic stellar population analysis, based on both full spectral fitting and line-index analysis, we have found that: 
\begin{itemize}
    \item[i)] the UCMGs cover a wide range of degree of relicness (DoR), from 0.05 to 0.88, tracing a wide variety in SFHs; 
    \item[ii)] the metallicity profiles are generally constant in time for extreme relics, but can vary quite dramatically for objects with a medium or low DoR;
    \item[iii)] based on the dependence of the DoR on fraction of stellar mass assembled by $z\sim2$, the time at which 75\% of the mass was in place, and the time of final assembly, we can divide the UCMGs into three DoR groups. Objects with $\text{DoR}\lesssim0.3$ are characterised by \Mfrac$<0.7$, a constant $t_{75}\sim 8$ Gyr,  and a $t_{\rm fin}$ as large as the Universe age at the redshift of each object. At intermediate DoR more than 60\% of the mass was formed at early cosmic time and hence the objects can be considered as relics. For $0.3\lesssim{\rm DoR}\lesssim0.6$, \Mfrac increases and $t_{75}$ decreases with increasing DoR but $t_{\rm fin}$ is still $\sim t_{\rm Uni}$. Finally, for $\text{DoR}>0.6$, the totality of the stellar mass was assembled less than 3 Gyr from the Big Bang (hence $t_{\rm fin}<<t_{\rm Uni}$). 
    This indicates that by looking at the three quantities combined into the DoR, one can truly distinguish between the variety of SFHs found among the UCMGs;
    \item[iv)] UCMGs with $\text{DoR}\gtrsim0.6$ are older by construction and are consistently metal richer, hitting the maximum value allowed by the SSP models with a very small scatter. They also have larger stellar velocity dispersion, smaller sSFRs, and larger [Mg/Fe] ratios. They are undoubtedly the most extreme relics of the Universe.
\end{itemize}

We confirmed 374 new ultra-compact objects at $0.01<z<0.3$ that have formed the majority ($\gtrsim70$\%) of their stellar mass during the first phase of the two-phase formation scenario. Of these, 81 were fully in place by $z\sim2$ and assembled the entire totality of their stars soon ($\lesssim3$ Gyr) after the Big Bang.  
The main conclusion of this paper is thus that selecting compact objects with a combination of large velocity dispersion values ($\sigma_{\star} \ge 250$\kms), super-solar metallicities ([M/H] $\ge 0.2$), old ages, and high [Mg/Fe] ratios is the most efficient way to find relics. 

Having assembled a statistically large sample of UCMGs with a variety of DoR will allow us, in the future, to compare this population with normal-sized galaxies of similar stellar masses. Indeed, one of the future plans of \EINSPIRE\ will be to compute the DoR in the innermost region of larger ETGs at similar redshifts, and compare the results we obtained for UCMGs so far.



\section*{Data Availability}
An online master catalogue presenting the stellar population results, as well as morphological and photometrical quantities for all 430 UCMGs is associated to this publication. This is publicly available for download from the \EINSPIRE\ website: \url{https://sites.google.com/inaf.it/chiara-spiniello/e-inspire}

\section*{Acknowledgements}
CS and CT acknowledge funding from the INAF PRIN-INAF 2020 program 1.05.01.85.11. 
AFM has received support from RYC2021-031099-I and PID2021-123313NA-I00 of MICIN/AEI/10.13039/501100011033/FEDER,UE, NextGenerationEU/PRT. CT acknowledges the INAF grant 2022 LEMON. GD acknowledges support by UKRI-STFC grants: ST/T003081/1 and ST/X001857/1. JPVB received the support of a fellowship from the "la Caixa" Foundation (ID 100010434). The fellowship code is LCF/BQ/DI23/11990084.



\bibliographystyle{mnras}
\bibliography{biblio_INSPIRE} 




\appendix

\section{Testing the impact of the fitting assumptions on the DoR}
\label{app:testing}

For each UCMG in the catalogue that we have assembled in Section~\ref{sec:data}, we have derived SFHs and therefore the DoR. This dimensionless number is very useful in quantifying the fraction of very old stars formed during the first phase of the two-phase formation scenario \citep{Oser+10, Naab+14, Huertas-Company+16}. However, we have also stressed in the main body of the paper that this quantity strongly depends on the choices one makes for the full-spectral fitting. Therefore, in this appendix, we provide a more detailed and technical description of the impact of our assumptions on the DoR (normalisation and distribution) and hence on the results presented in Section~\ref{sec:results}.

\subsection{Changing the threshold for the final assembly time}
\begin{figure}
    \centering \includegraphics[width=\columnwidth]{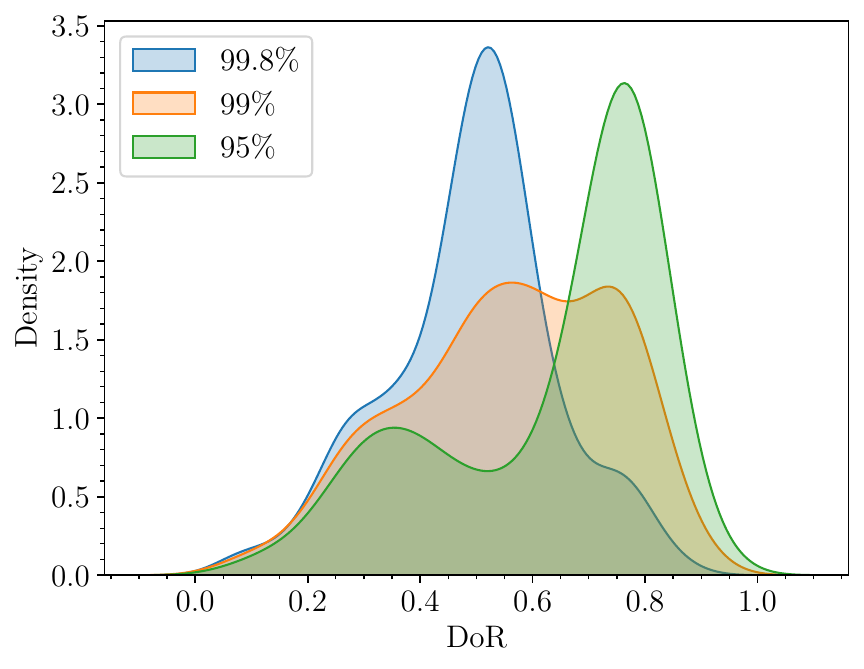}
\caption{A comparison of how the DoR distributions changes for different choices of threshold when defining $t_{\rm fin}$. In blue, the distribution for when this threshold is 99.8\% of the stellar mass in place (as in this paper). In orange, the distribution for when this threshold is 99\%. In green, the distribution when this threshold is 95\%. This plot is produced by smoothing the distribution with a Gaussian kernel to improve visibility. }
    \label{fig:dor_tfin}
\end{figure}

Among the different quantities used to compute the DoR, $t_{\rm fin}$ is that with the largest uncertainty (as noted in Section~\ref{sec:dor}). We believe that this is because the spectral fit on the optical spectra is very sensitive to a sub-percentage fraction of very recent star formation.  Especially when bootstrapping, it is likely that \ppxf\ picks a few SSP models with very young ages, hence increasing the final time of assembly. This is especially evident when requiring a very conservative threshold of $99.8$\% of the stellar mass being assembled, as we do in the main body of the paper. Hence, here we test the impact on the overall DoR distribution by lowering this threshold to 99 and to 95 per cent. Figure~\ref{fig:dor_tfin} shows the current DoR, as obtained in the main text, in blue and the distributions with lower thresholds in orange and green respectively. As expected, reducing the threshold of stars being assembled pushes the distribution of DoRs towards the higher values. Interestingly, the 99\% threshold flattens the peak at $\sim0.5$, broadening the distribution across a larger range of DoR. Lowering the threshold to 95\% instead produces a peak at much higher DoR, with 291 objects classified as extreme relics ($\text{DoR}>0.6$).

\subsection{Using mean or median quantities when computing DoR}
\begin{figure}
    \centering  \includegraphics[width=\columnwidth]{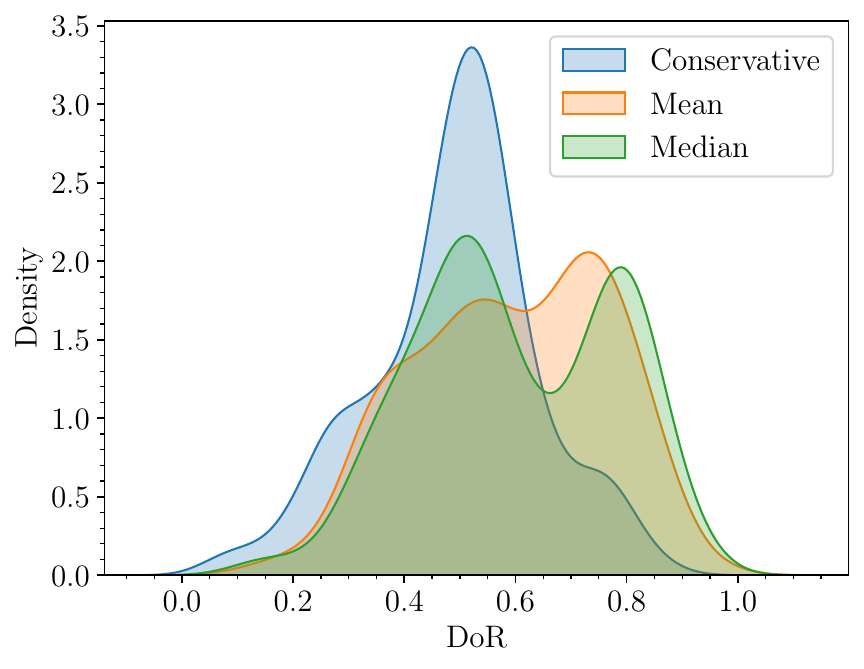}
    \caption{A comparison of the difference in DoR distributions when we use different estimates of \Mfrac, $t_{75}$, and $t_{\rm fin}$. Here we plot the distributions for the most conservative, mean, and median estimates of each quantity. This plot is produced by smoothing the distribution with a Gaussian kernel to improve visibility.}
    \label{fig:dor_est}
\end{figure}

In the computation of the DoR, we have used the most conservative estimates of the three single parameters: the minimum of \Mfrac, the maximum $t_{75}$ and $t_{\rm fin}$. Clearly, this approach returns the minimum DoR for each object. 
Figure~\ref{fig:dor_est} shows how the distribution would change if we used mean or median quantities instead. 
Predictably, using mean quantities shifts the overall distribution of DoRs towards much higher values. As a result, the number of extreme relics goes from 81 to 216 and the number of non-relics dramatically reduces from 56 to 16. The number of objects with $0.3\lesssim\text{DoR}\lesssim0.6$ goes from 293 to 198.
Using median quantities instead produces two different peaks, one identical to the original distribution derived with the most conservative parameters, and the other following the mean distribution. We also note that at very low DoR, the three distributions are similar.

\subsection{Computing light-weighted ages and metallicities}
\begin{figure}
    \centering    \includegraphics[width=\columnwidth]{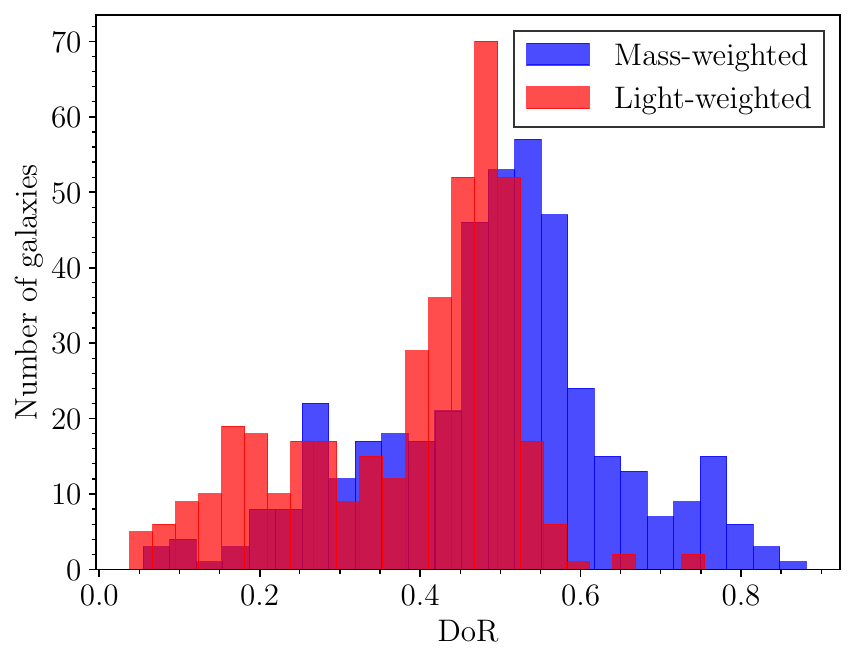}
    \caption{A comparison of the DoR distributions computed using mass-weighted population parameters versus light-weighted population parameters.}
    \label{fig:light_mass}
\end{figure}

Following previous \INSPIRE\ publications, we have inferred mass-weighted ages and metallicities from the full-spectral fitting. This choice is motivated by the fact that mass-weighted quantities offer a direct probe of the integrated SFH and MEH. 
Since young stars outshine evolved ones, mass-weighted ages are always older than light-weighted ones.
Furthermore, light-weighted ages and metallicities tend to artificially strengthen the observed trends with other parameters (e.g. velocity dispersion, \citealt{Trager09}). 
In Figure~\ref{fig:light_mass} we compare the DoR histograms obtained when computing mass-weighted ages (blue) and light-weighted ones (red). As expected, the overall effect is that the number of non-relics (DoR$<0.3$) is larger in the latter case.  

\subsection{Removing flagged objects}
\begin{figure}
    \centering    \includegraphics[width=\columnwidth]{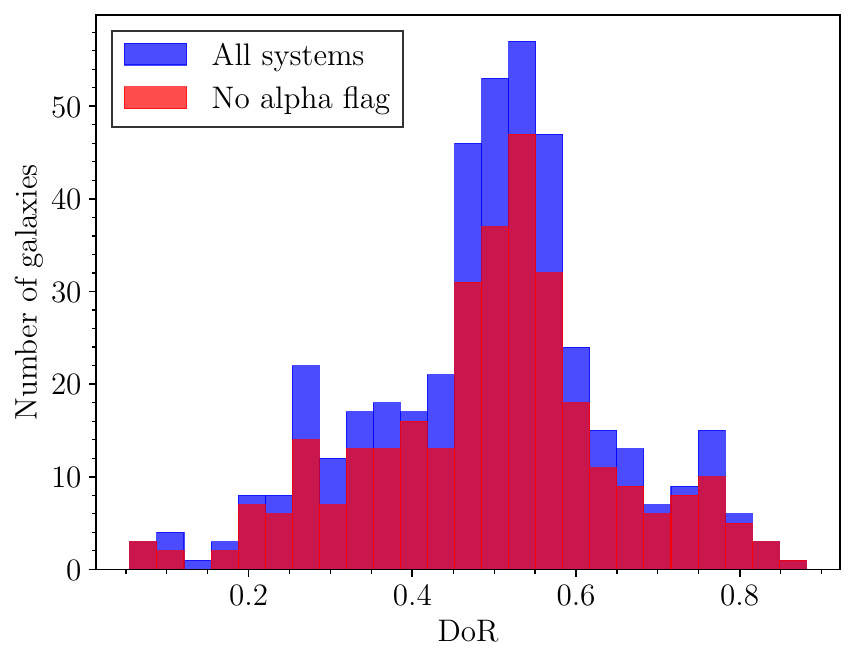}
    \caption{A comparison of the DoR distributions for the full set of UCMGs and excluding objects which were flagged (see Section~\ref{sec:index}).}
    \label{fig:dor_alpha}
\end{figure}

In Section~\ref{sec:index}, we have used the Mg$_b$-$\langle$Fe$\rangle$ index--index plot to infer the [Mg/Fe] abundances for the entire sample of 430 UCMGs. We have used the MILES SSP models to build a grid with \afe\ varying from 0 to 0.4 in steps of 0.1 on which we overplotted the systems. However, 116 objects fall outside the model grid and hence have been flagged. For these, we adopted the most conservative approach and ran \ppxf\ using the two models with extreme \afe\ values (0.0 and 0.4). 
In the \citetalias{Spiniello20_Pilot}, we have shown that as \afe\ is increased, the ages generally become older and the SFHs become slightly more peaked. Hence, using the maximum range of \afe\ values to derive SFHs for the flagged objects produces the largest uncertainties and gives the most conservative result. Here, we confirm that this does not have an impact on the results we presented in the main paper. In fact, Figure~\ref{fig:dor_alpha} shows that the distribution of the DoR when excluding the flagged objects is very similar to the one where they are included. 

\subsection{Fitting with `unsafe' SSP models}
\begin{figure}
    \centering   \includegraphics[width=\columnwidth]{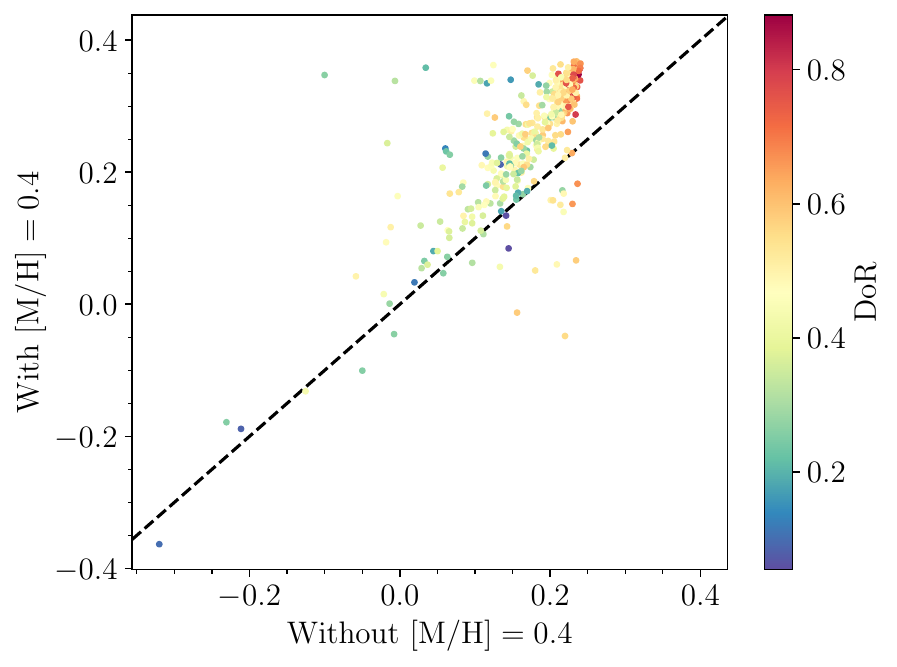}
    \caption{A comparison of how the computed metallicity changes when we include the $\text{[M/H]}=0.40$ dex models. The red dashed line indicates the 1-to-1 relation. The points are colour-coded by their DoR calculated without the $\text{[M/H]}=0.40$ dex models.}
    \label{fig:z_scatter}
\end{figure}

\begin{figure}
    \centering    \includegraphics[width=\columnwidth]{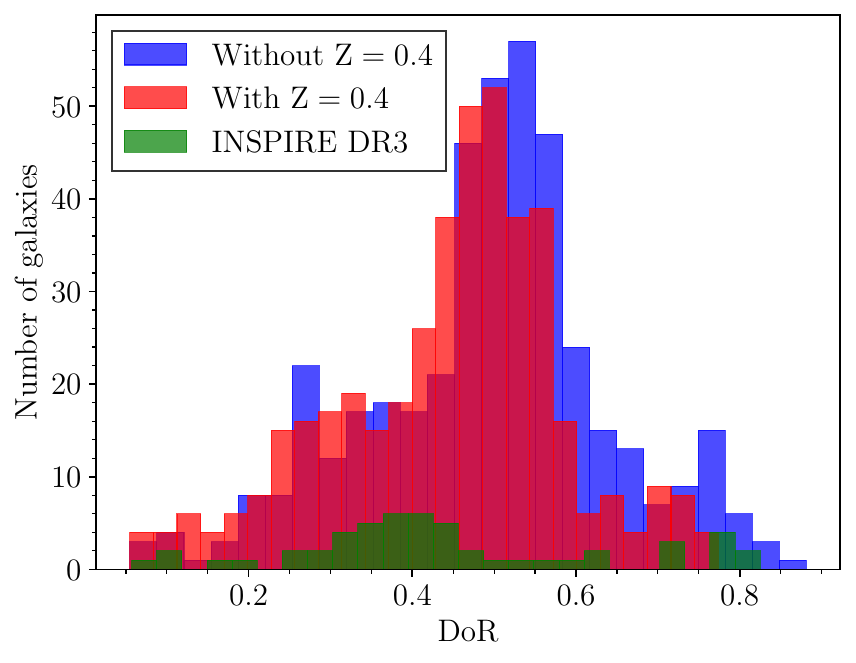}
    \caption{A comparison of the DoR distributions when limiting ourselves to SSP models in the `safe ranges' (i.e. not including models with $\text{[M/H]}=0.40$ dex) and when we use all available models (including $\text{[M/H]}=0.40$ dex). The DoR distribution from \citetalias{Spiniello24} (which uses $\text{[M/H]}=0.40$ dex models) is included for comparison.}
    \label{fig:dor_z}
\end{figure}

As noted in Section~\ref{sec:stel_pop_ppxf}, we limit ourselves to SSP models in the `safe ranges', i.e. those with metallicities up to $\text{[M/H]}=0.26$ dex. This was strongly advised by the author of the models (private communication). However, our results suggest that relics tend to have super-solar metallicities, with a significant number reaching the highest metallicity allowed by the models. Here we thus investigate the impact of including these `unsafe' models. 
In Figure~\ref{fig:z_scatter} we plot the metallicities estimated without the $\text{[M/H]}=0.40$ dex SSP models (main text) versus these estimated including them to show that this causes a shift towards higher metallicity for all systems in our sample, despite their DoR. 
Figure~\ref{fig:dor_z} shows the DoR distribution shifts slightly towards lower values when $\text{[M/H]}=0.40$ dex is included, suggesting that the effect of this change is to decrease the ages of these galaxies. This is unsurprising due to the age-metallicity degeneracy \citep{Worthey+94, Worthey+99}. We also note that the distribution including $\text{[M/H]}=0.40$ dex is closer to that of the main paper compared to that of \citetalias{Spiniello24}, where indeed the 'unsafe' models were included in the \ppxf\ fit.


\section{Comparison between size estimates from SDSS and from KiDS}
\label{app:size_compare}

In this appendix, we compare the effective radii calculated in this paper (as described in Section~\ref{sec:ucmg}) with the effective radii measured from KiDS data \citep{Tortora+18_UCMGs, Scognamiglio20}. 
This is particularly crucial as estimating sizes of UCMGs from ground-based, seeing-limited imaging is very challenging. We point the readers to the Appendix B of \citet{Tortora+18_UCMGs}, which performs a detailed analysis on the expected systematics and statistical uncertainties on the effective radii estimates. 

\begin{figure}
    \centering    \includegraphics[width=\columnwidth]{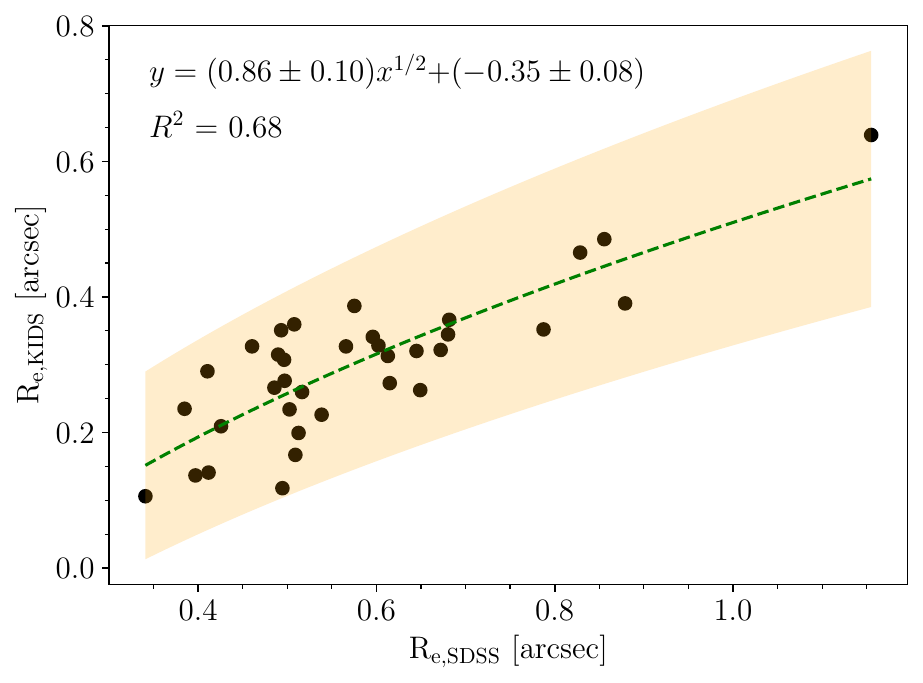}
 \caption{A comparison of the effective radii measured by SDSS and KiDS for a selection of systems in the \INSPIRE\ sample. This is fitted with a quadratic (the green dashed line), of which the equation (with errors) is shown. The $R^2$ value for this fit is also included.}
\label{fig:kids_compare}
\end{figure}

To compare the size estimates from SDSS and KiDS, we consider systems from the \INSPIRE\ sample that are within SDSS DR18 \citep{Almeida+23_SDSS}, of which there are 36 (out of 52). For these systems, we calculate their effective radii as measured by SDSS using the same procedure as Section~\ref{sec:ucmg}. This measure of \Reff\ is compared with that measured by KiDS in Figure~\ref{fig:kids_compare}.

From this figure, we clearly see that the effective radii computed by SDSS are systematically larger than those inferred from KiDS. We speculate that this is caused by the worse spatial resolution (larger pixel size) and PSF FWHM of SDSS, which makes it harder to estimate precise effective radii, especially for these incredibly small objects. 

We find that this scatter is fitted well by a non-linear equation with $R^2=0.68$ with $\mathrm{R}_{\mathrm{e, SDSS}}$ scaling as $\sqrt{\mathrm{R}_{\mathrm{e, KiDS}}}$. 
This indicates that the relation between the radii is quadratic in nature, which we suggest might be due to the PSF as the radial linear size goes with the 2D shape of the PSF, which relates to its area. 
Thus, we fit a quadratic relation to the points, deriving the following equation: 

\begin{equation}
\label{eq:size}
\mathrm{R}_{\mathrm{e, SDSS}} = (0.86\pm0.10)    \mathrm{R}^{1/2}_{\mathrm{e, KiDS}} + (-0.35\pm0.08).
\end{equation}

\begin{figure}
    \centering    \includegraphics[width=\columnwidth]{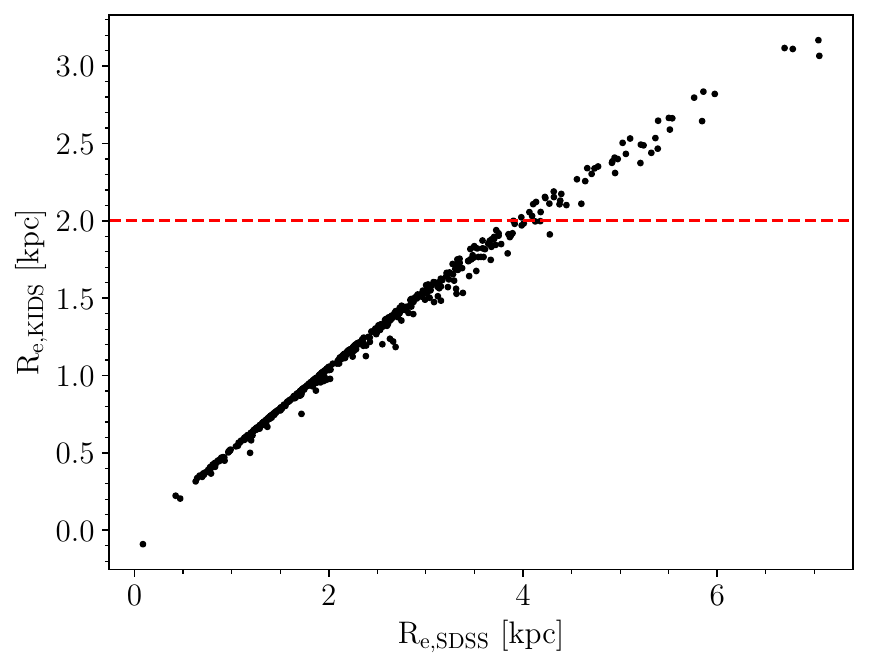}
 \caption{Sizes computed from SDSS imaging plotted against sizes corrected as they will be measured from KiDS higher resolution images.}
\label{fig:radii_compare}
\end{figure}

We then use this equation to correct the sizes we derive for all the objects selected in Section~\ref{sec:ucmg}, and hence to assemble the final sample of 430 ultra-compact galaxies.  Figure~\ref{fig:radii_compare} shows the two estimates of the effective radii one against each other. The horizontal dashed red line highlights the 2 kpc threshold that has been used, as in previous \INSPIRE\ papers, to select 430 UCMGs. Given the worse resolution of SDSS imaging, this same threshold would correspond to sizes in SDSS of $\lesssim4$ kpc. Based on this correction, we finally select 446/495 objects (of which 430 are UCMGs, see main text for more details). 

As already pointed out in \citetalias{Spiniello24}, the only way to precisely estimate structural parameters for UCMGs is to obtain AO supported or space observations.

\bsp	
\label{lastpage}
\end{document}